\documentclass[%
 reprint,
superscriptaddress,
 amsmath,amssymb,
 aps,
pre
]{revtex4-1}

\usepackage{xcolor}
\usepackage{hyperref}
\usepackage{graphicx}
\usepackage{dcolumn}
\usepackage{bm}
\usepackage{booktabs}
\usepackage{hyperref}
\usepackage{subfigure}

\DeclareMathOperator{\csch}{csch}

\begin{document}

\title{Emergence of an Ising critical regime in the clustering of 1D soft matter revealed through string variables}

\author{F. Mambretti}
\affiliation{
 Universit\`a degli Studi di Milano, Dipartimento di Fisica, via Celoria 16, 20133 (Milano), Italy
}
\author{S.Molinelli}
\affiliation{
 Universit\`a degli Studi di Milano, Dipartimento di Fisica, via Celoria 16, 20133 (Milano), Italy
}
\author{D. Pini}
\affiliation{
 Universit\`a degli Studi di Milano, Dipartimento di Fisica, via Celoria 16, 20133 (Milano), Italy
}
\author{G. Bertaina}
\affiliation{
 Istituto Nazionale di Ricerca Metrologica, Strada delle Cacce 91, 10135 (Torino), Italy
}
\affiliation{
 Universit\`a degli Studi di Milano, Dipartimento di Fisica, via Celoria 16, 20133 (Milano), Italy
}
\author{D.E. Galli}
\affiliation{
 Universit\`a degli Studi di Milano, Dipartimento di Fisica, via Celoria 16, 20133 (Milano), Italy
}

\begin{abstract}
Soft matter systems are renowned for being able to display complex emerging phenomena such as clustering phases.
Recently, a surprising quantum phase transition has been revealed in a one--dimensional (1D) system composed of bosons interacting via a pairwise soft potential in the continuum.
It was shown that the spatial coordinates undergoing two-particle clustering could be mapped into quantum spin variables of a 1D transverse Ising model.
In this work we investigate the manifestation of an analogous critical phenomenon in 1D classical fluids of soft particles in the continuum. In particular, we study the low--temperature behavior of three different classical models of 1D soft matter, whose inter--particle interactions allow for clustering.
The same string variables highlight that, at the commensurate density for the two--particle cluster phase, the peculiar pairing of neighboring soft particles can be nontrivially mapped onto a 1D discrete classical Ising model. We also observe a related phenomenon, namely the presence of an anomalous peak in the low--temperature specific heat, thus indicating the emergence of Schottky phenomenology in a non-magnetic fluid.
\end{abstract}

\maketitle

\section{Introduction}\label{sec:intro}

Soft matter systems are characterized by an interaction potential which allows for a finite probability 
for two or more particles to overlap. 
Thanks to this feature, they may show a tendency to self--organize into mesoscopic structures, 
and the properties and the interactions of these structures may determine the macroscopic 
behavior of the system.
Recently, the interest in soft potentials has grown: classically they were
found to describe the behavior of compenetrating polymers, molecules\cite{ref_uno} and liquid crystals\cite{ref_due},
while recent applications in quantum physics were devised in the field of ultracold gases\cite{ref_tre} and in the search 
for exotic super-solid phases\cite{ref_quattro,ref_cinque}.
In the last few decades the key features of the phase diagram of many quantum and classical soft potentials were studied 
in three and lower dimensions, mainly with computational methods\cite{ref_quattro,ref_cinque,ref_sei,ref_sette,ref_otto,ref_nove,ref_dieci}. 
It was found that not every soft potential shows the tendency to form aggregates
at high densities - i.e ``clustering''\cite{ref_undici} - and that the key feature to observe
such behavior is a negative global minimum of the Fourier transform of the potential\cite{ref_otto}. 
For such clustering potentials, the average number of particles per cluster increases with density, 
and homogeneous $n$--cluster phases (i.e. with $n$ particles per cluster) appear at specific commensurate densities.
One dimensional systems, in this respect, are peculiar, because finite-temperature phase transitions are prevented in systems with either hard-core interactions or discrete (spin) degrees of freedom by the van Hove theorem \cite{vanhove:1950} in classical mechanics and Mermin-Wagner theorem \cite{mermin:1966} in quantum mechanics. There has been debate about whether such theorems can be extended to soft potentials \cite{Cuesta2004, Fantoni2010, Acedo2004, Speranza2011}.
Recently, a new zero--temperature quantum phase transition (QPT) has been discovered in the clustering of a particular quantum soft matter 
system in one dimension\cite{ref_sei}. By means of quantum Monte Carlo simulations it has been observed that, at the commensurate density for the two--particle cluster phase, the soft system has a secondary excitation mode which is gapless 
only at the transition point. Further analysis showed that this soft excitation can be mapped via string variables
onto an effective one--dimensional (1D) quantum transverse Ising model, finding that the tendency of the soft system to form clusters is 
the crucial feature for this mode to appear. 
This fascinating fact opens the interesting possibility that these ``magnetic-like'' excitations could be a common 
feature of all the clustering potentials, for both classical and quantum systems.

In this work, our aim is to find evidence of the presence of the same kind of excitations in classical clustering 
fluids in one dimension, using stochastic simulation techniques. 
By considering three different soft models, we find that at very low temperatures, at the specific commensurate densities, 
critical two--particle cluster regimes appear. 
Under these thermodynamic conditions, via the same string variables introduced for the quantum transition,
we observe the emergence of a critical regime related to a one--dimensional Ising model of pseudo--spins, consistent with a zero--temperature critical point.
Moreover, the specific heat of the soft system is characterized by a typical phonon contribution at low temperature, while it shows an anomalous peak at intermediate temperatures, similarly to the 1D classical Ising model. 
The physical properties of the equivalent pseudo--spin system are also measurable as a function of the temperature $T$. We show here that the values of the 
susceptibility, energy, specific heat and spin--spin correlation functions approach the theoretical Ising curves in the limit of zero temperature. This leads to a behavior which is model--independent; moreover, at fixed pair potential, from all the physical properties a consistent estimate of the coupling constant $J$ of the Ising Hamiltonian can be extracted. We suggest here that the $J$ parameters calculated from the various physical quantities tend towards a unique $T\to 0$ limit value. Remarkably, a consistent value can be also deduced from a study of the energy of the ``defected" configurations of the system, computed via Simulated Annealing (SA)\cite{kirkpatrick:1983}. At relatively higher temperatures, we find that the anomalous specific heat peak is consistently much higher than for the non--clustering Gaussian potential and has surprisingly universal features, similarly to solid--state magnetic systems displaying a Schottky anomaly.

The structure of the paper is as follows. In Sec.~\ref{sec:model}, the three pair potentials investigated and the details of their clustering properties are presented. Sec.~\ref{sec:results} contains the description of the thermodynamic physical properties found for the soft particles, via canonical Monte Carlo simulations. In Sec.~\ref{sec:mapping}, we introduce the nontrivial mapping between the soft--particle spatial degrees of freedom and a system of Ising pseudo--spins on a lattice. We anticipate that this procedure sends many degrees of freedom onto few ones, whose properties we aim to characterize. In this regard, we show the Ising-like properties of the pseudo--spin system, highlighting the intriguing agreement between the simulated data and the theoretical Ising curves in the limit for $T \rightarrow 0$. Sec.~\ref{sec:jeff} reports the investigation of the temperature dependence of the coupling constant $J$ of the Ising Hamiltonian, pursued by a statistical analysis of the soft configurations and by the SA optimization method. This is followed by Sec.~\ref{sec:scaling}, in which we discuss the universal scaling of the pseudo--spin observables and the relation between the soft--particle and pseudo--spin specific heats. We show that the energy fluctuations (i.e. the specific heat) of the pseudo--spin variables are intimately and surprisingly related to the corresponding properties of the original system.  Our conclusions are presented in Sec.~\ref{sec:conclusions}.

\section{Physical models and simulation method}\label{sec:model} 

\begin{figure}[tbp]
  \includegraphics[width=\columnwidth]{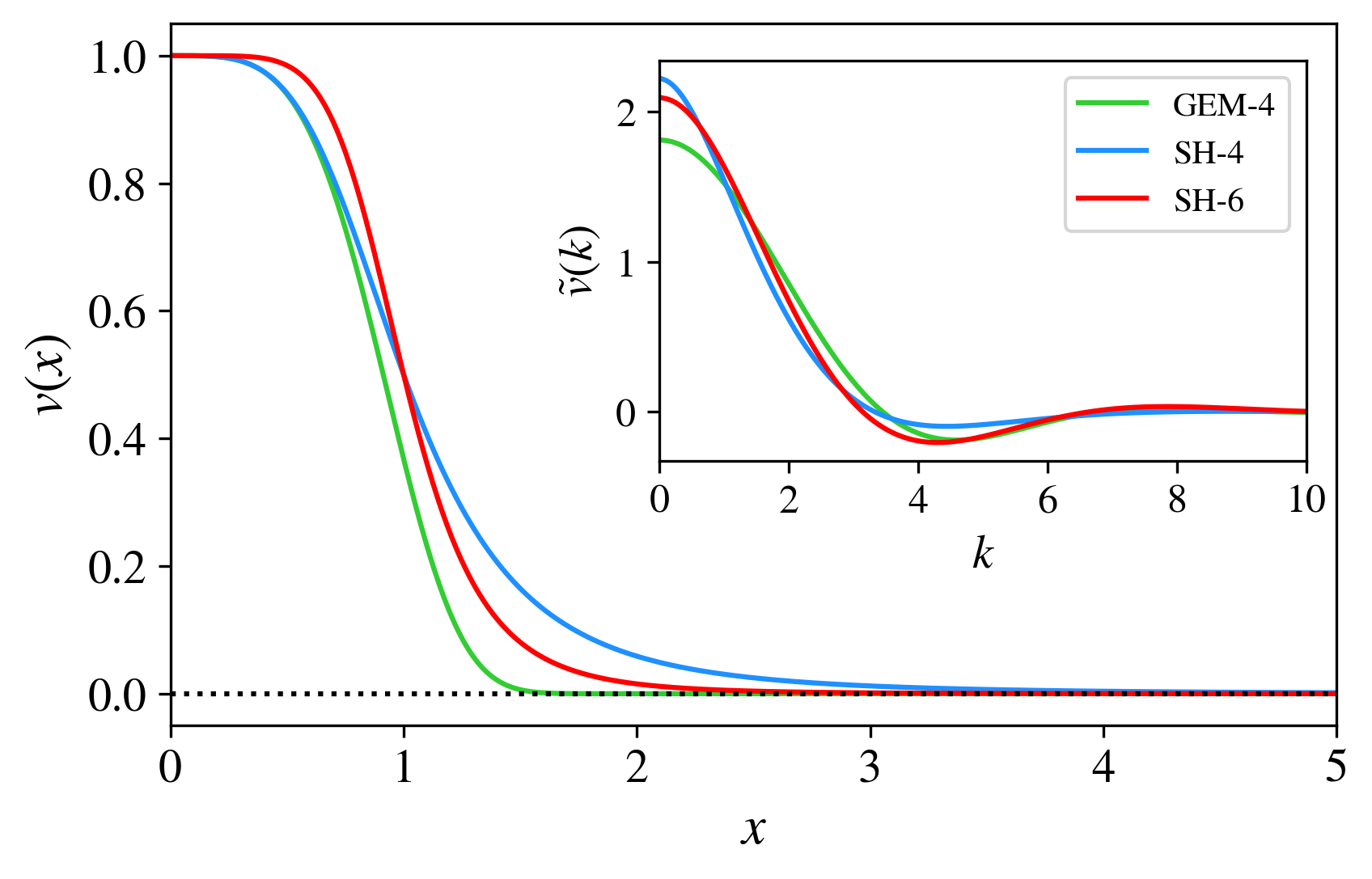}
  \caption{Inter--particle interaction potentials $v(x)$. In the inset their Fourier transforms, $\tilde{v}(k)$ are drawn. They all display a negative minimum, thus satisfying Likos criterion for clustering. See also Tab.~\ref{tab:density}.}
  \label{fig:fourier}
\end{figure}

In this work, the results for three distinct soft--core pair potentials are reported. We restrict our investigation to 1D systems of particles characterized by limited, positive, purely repulsive and short-range interactions.
In particular, we consider the \textit{Generalized Exponential Model of order four} (also known as GEM--4 and extensively studied in\cite{prestipino2:2014, prestipino:2015}) pair interaction, which has the following functional form:
\begin{equation}
v(r)= U\:e^{-(\frac{r}{\sigma})^4}
\label{eqn:gem4}
\end{equation}
and the \textit{Shoulder--4} (SH--4) and \textit{Shoulder--6} (SH--6) potentials, where a \textit{Shoulder--}$m$ pair interaction is described by:
\begin{equation}
v(r)=\frac{U}{1+(\frac{r}{\sigma})^m}
\label{eqn:shoulder}
\end{equation}
In the previous equations, $r$ is the inter--particle distance, $\sigma$ is the characteristic length scale and $U$ represents the interaction intensity.
All these potentials satisfy Likos criterion concerning the presence of a negative part of their Fourier transform\cite{ref_otto, ref_undici, likos:2007} and, therefore, admit high--density $n$--particle clustering phenomena at low temperatures. These pair interactions are said to belong to the $Q^{\pm}$ class.
In the following, energy is naturally measured in units of $U$ and the distances are measured in units of $\sigma$. Therefore, equations \eqref{eqn:gem4} and \eqref{eqn:shoulder} become respectively: \begin{equation}
v(x)=e^{-x^4}
\end{equation}
and \begin{equation}
v(x)=\frac{1}{1+x^m}
\end{equation}
where we have set $x=r/\sigma$. As a consequence, physical wavevectors $q$ will be measured in units of $\sigma^{-1}$, i.e. $k=q\,\sigma$. The same applies to number densities $\rho = N \, \sigma / L $, in a system with $N$ particles in a box of length $L$, always in periodic boundary conditions (PBC).
Reduced temperatures $t=k_B\,T/U$ are also used throughout the article. 
In Fig.~\ref{fig:fourier} we display the three considered potentials and their Fourier transforms.

Table~\ref{tab:density} displays the values of the wavenumber $k_{min}$ corresponding to the minimum value of the Fourier transform of our potentials and the corresponding optimal reduced densities for the formation of two--particle clusters, where \begin{equation}
\rho_{(n)} = \frac{n\:k_{\text{min}}}{2\pi}
\end{equation}
and, therefore, $\rho_{(2)}=k_{min}/\pi$.

\begin{table}[tbp]
\begin{ruledtabular}
\begin{tabular}{lll}
Pair potential & $k_{min}$ & $\rho_{(2)}$\\
\hline
GEM--4 & 4.59180 & 1.46165\\
SH--4 & 4.44289 & 1.41422\\
SH--6 & 4.29952 & 1.36857\\
\end{tabular}
\end{ruledtabular}
	\caption{$k_{min}$ and $\rho_{(2)}$ for the three pair potentials studied}
	\label{tab:density}
\end{table}

In our calculations, each of the three systems is studied at its own commensurate numerical density $\rho_{(2)}$, for the two--particle cluster phase. The study of this system at non--commensurate densities is clearly a very interesting topic, which has already been explored \cite{prestipino:2015,neuhaus:2011}, and is outside the scope of this work. Low--temperature thermodynamic properties are computed via canonical Monte Carlo (MC) simulations by using the Metropolis algorithm to sample equilibrium configuration of the $N$ particles.

\section{Results for the soft--particle system}\label{sec:results}

\begin{figure}[tbp]
  \includegraphics[width=\columnwidth]{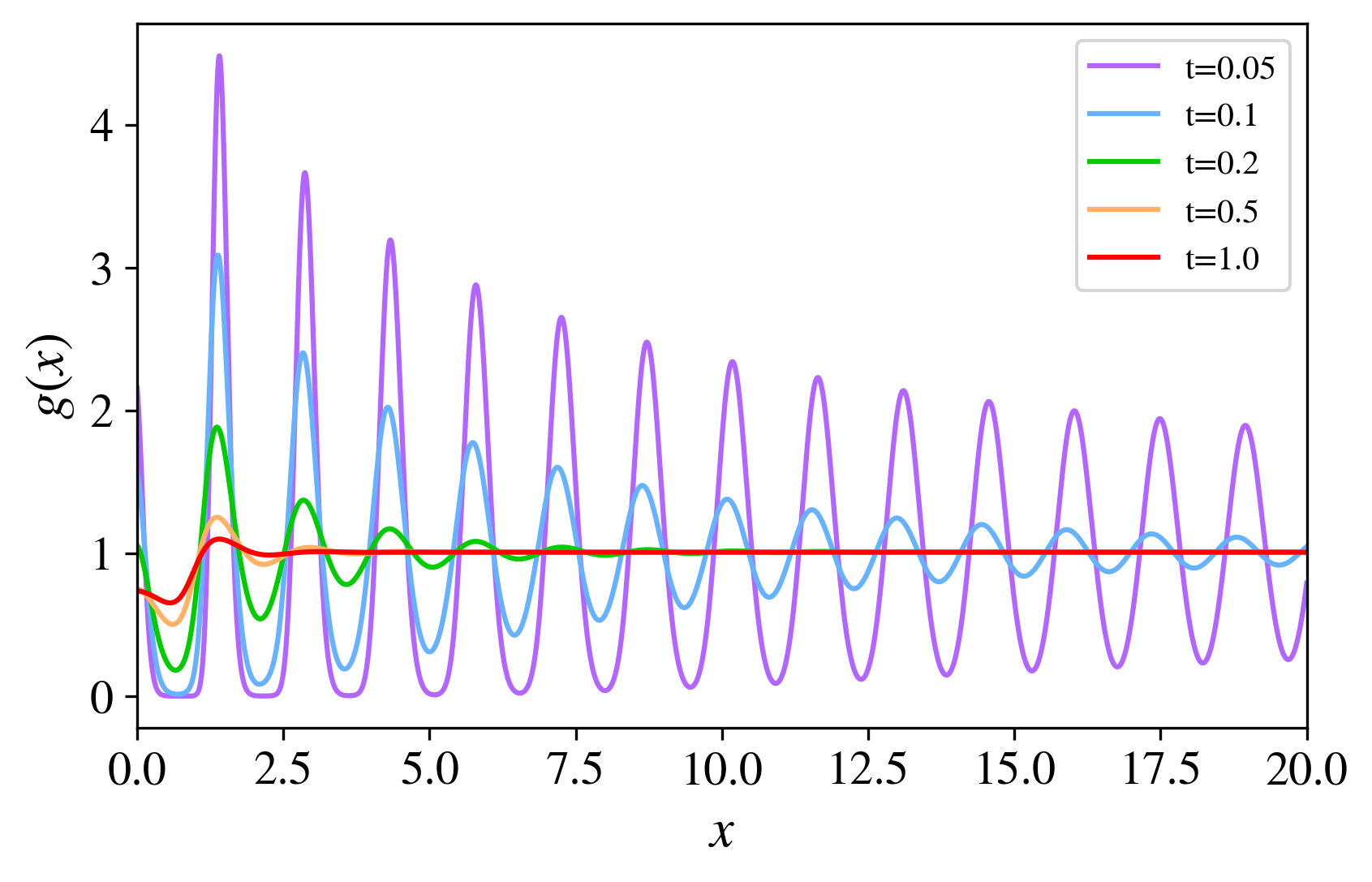}
  \caption{Temperature dependence of the pair distribution functions $g(x)$ for the SH--6 potential. Errorbars are smaller than the line width.}
  \label{fig:gofr}
\end{figure}

In this section, we show the results of the MC simulations for the thermodynamic quantities calculated from the low--temperature equilibrium configurations of the 1D soft particles system. 
This first part of the paper is similar to the analysis performed by Speranza \emph{et al.} \cite{Speranza2011} on the Gaussian core model. They investigated the properties of the 1D GEM--2 soft fluid, finding anomalies in the specific heat and determining the structural properties of the system via pair correlation functions and static structure factors. In particular, they suggested the presence of traces of the ordered arrangements found in clustered crystals, despite the absence of a true solid--liquid phase transition. The results reported in the following are analogous, but are referred to genuine clustering potentials, at variance with the GEM--2.
Van Hove's theorem does not hold, due to the lack of any hard--core interaction in our model, as discussed e.g. in \cite{Cuesta2004}, despite being in 1D. Previous studies gave strong evidence that the penetrable spheres and the penetrable square well models do not display a thermal phase transition in 1D \cite{Fantoni2010}; conversely, Acedo \emph{et al.} \cite{Acedo2004} identified the appearance of a crystalline one--dimensional phase for a step (discontinuous) potential, which is not our case. As a consequence, it is not known \emph{a priori} if a critical finite--temperature phase transition does exist for our models. Our results, in agreement with previous studies \cite{Speranza2011}, show that a phase transition is eventually expected only in the limit $t\rightarrow0$.
This is immediately apparent in the features of the two--particle distribution function $g(x)$ upon decreasing $t$, where $g(x)=\dfrac{\rho^{(2)}(x_1,\dots,x_N)}{\prod_{i=1}^{N} \rho^{(1)}(x_i)}$ and the $p$--particles density is equal to $\rho^{(p)}=\dfrac{N!}{(N-p)!} \displaystyle \int \rho(x_1, \dots, x_N) dx_{p+1} \dots dx_N$. 
The temperature dependence of $g(x)$ for the SH--6 potential is shown in Fig.~\ref{fig:gofr} for some relevant $t$ values. Results for the other potentials are essentially equivalent. This fluid was simulated at density $\rho_{(2)}=1.36857$; in these conditions, a perfect single--particle crystal would host a particle about every $\Delta x = 0.73069$. 
A key observation is that $g(x)$, by lowering the temperature, shows the formation of oscillations, damped with increasing distance $x$, at a spatial separation $2\Delta x$. Moreover, the formation of a peak in the origin can be observed, whose height grows while approaching $t=0$.
These features clearly highlight the tendency of the system to form clusters composed of two particles. The growing height of the peaks of the oscillations in $g(x)$ indicates the emergence of a quasi-long-range order, due to the increasing regularity of the spatial arrangement of the clustering particles. For much higher temperatures, this order is lost and $g(x)$ displays some structure only at small distances.

\begin{figure}[btp]
  \includegraphics[width=\columnwidth]{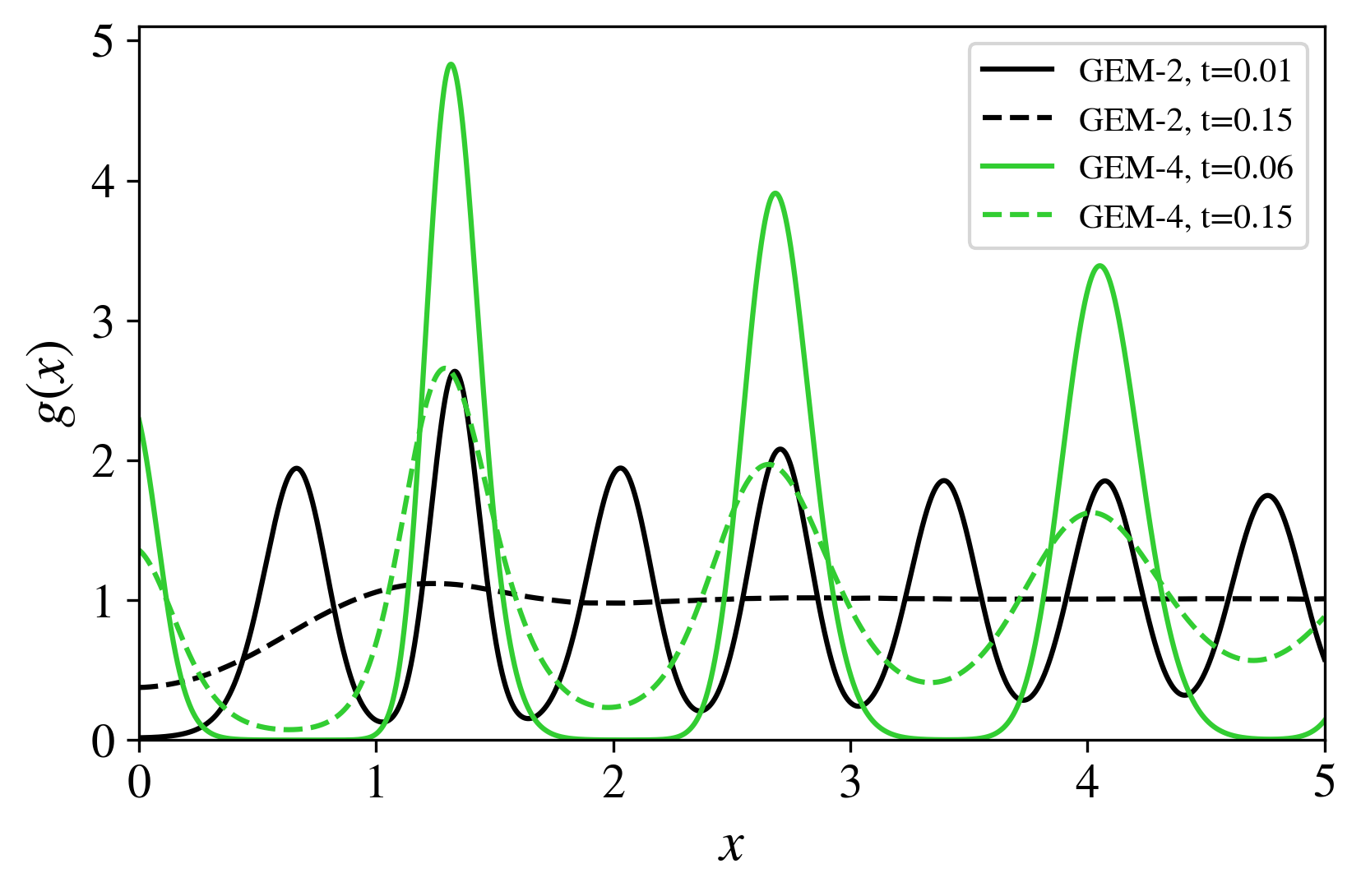}
  \caption{Comparison between the $g(x)$ of GEM--2 and GEM--4 potentials at various temperatures, computed at reduced density 1.46165. Errorbars are smaller than the line width.}
  \label{fig:gem2vsgem4_gofr}
\end{figure}

To further remark this behavior of the clustering potentials, in Fig.~\ref{fig:gem2vsgem4_gofr} we compare $g(x)$ for two systems of particles interacting via either the GEM--2 (i.e., Gaussian) or the GEM--4 potential, both simulated at reduced density 1.46165. We considered two simulations at different $t$ for each system, so as to be able to compare the properties at low and at high temperature of a clustering and a non--clustering system. The GEM--2 tends to show ordering at much lower temperatures~\cite{lang:2000}; for this reason, in Fig.~\ref{fig:gem2vsgem4_gofr} we compared a GEM--2 $g(x)$ at $t=0.01$ to the $g(x)$ for the  GEM--4 simulated at $t=0.06$.

The GEM--2 system displays a minimum for the occupation probability in $x=0$ (even at low temperature), while the GEM--4 clearly always presents a maximum in the pair distribution function at the origin. At low temperatures, the oscillations of $g(x)$ for the GEM--2 case display a periodic spacing which is one half that of the GEM--4. Therefore, also the particles interacting via the GEM--2 potential have some tendency to solid ordering, but they do not show any clustering tendency.
We observe a striking feature: the $g(x)$ peaks of the GEM--2 corresponding to the clustering peaks of the GEM--4 are slightly higher than the other peaks of the GEM--2. Interestingly, as already observed in \cite{Speranza2011} a non--clustering potential such as the GEM--2 displays, nonetheless, a slightly prevalent spatial ordering in correspondence of the peak positions characteristic of clustering phases. This is remarkable since the GEM--2 does not have any negative component in its Fourier transform.
By comparing the $g(x)$ for the two potentials at $t=0.15$ (dashed lines in the figure), we note that the particles interacting via the GEM--4 already present a partial order and a finite occupation in $x=0$, while the GEM--2 is still mostly disordered.

\begin{figure}[tbp]
  \includegraphics[width=\columnwidth]{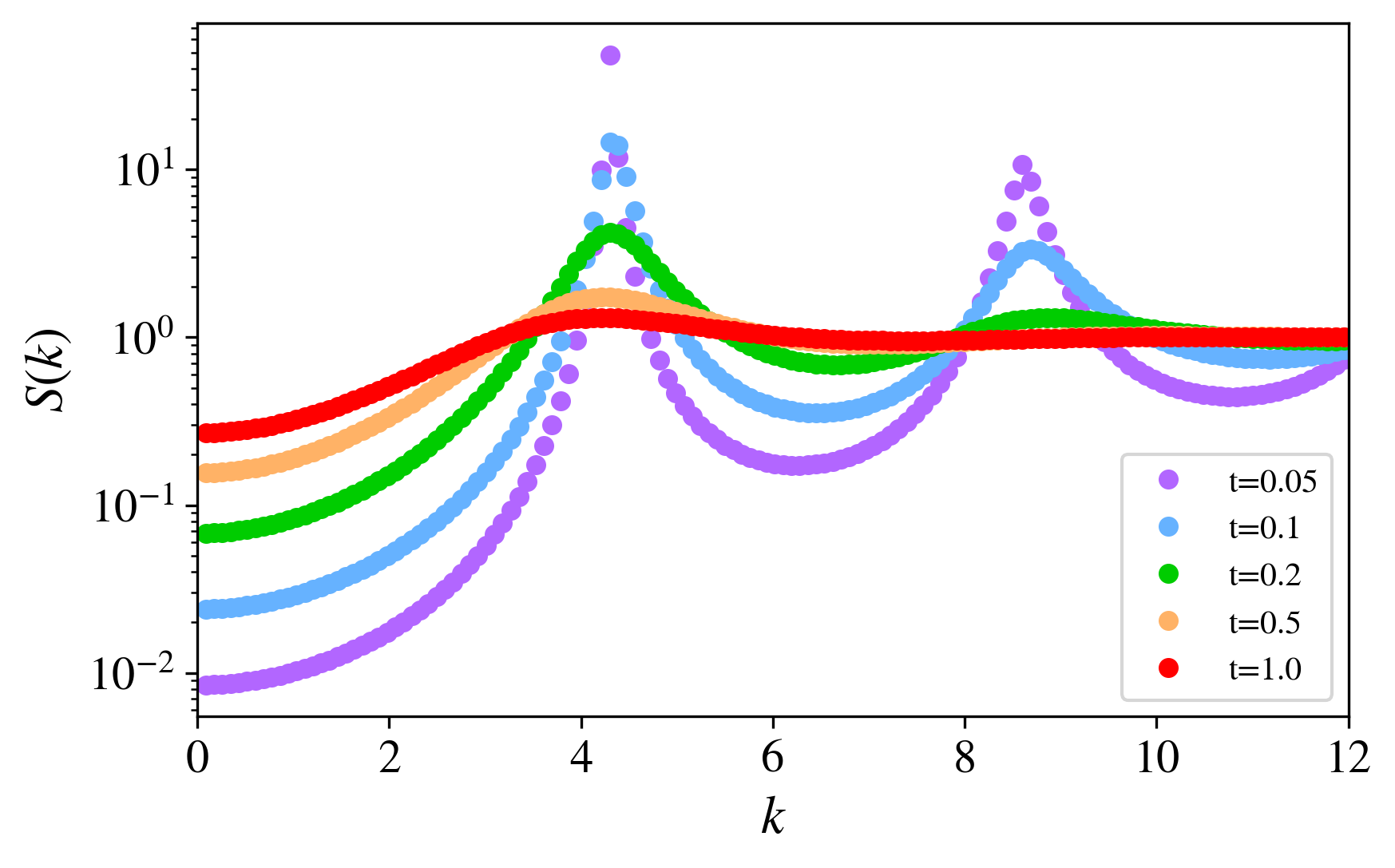}
  \caption{Temperature dependence of the static structure factor $S(k)$ for the SH--6 potential. Note the logarithmic scale for $S(k)$. Errorbars are smaller than the line width.}
  \label{fig:sofq}
\end{figure}

\begin{figure}[btp]
  \includegraphics[width=\columnwidth]{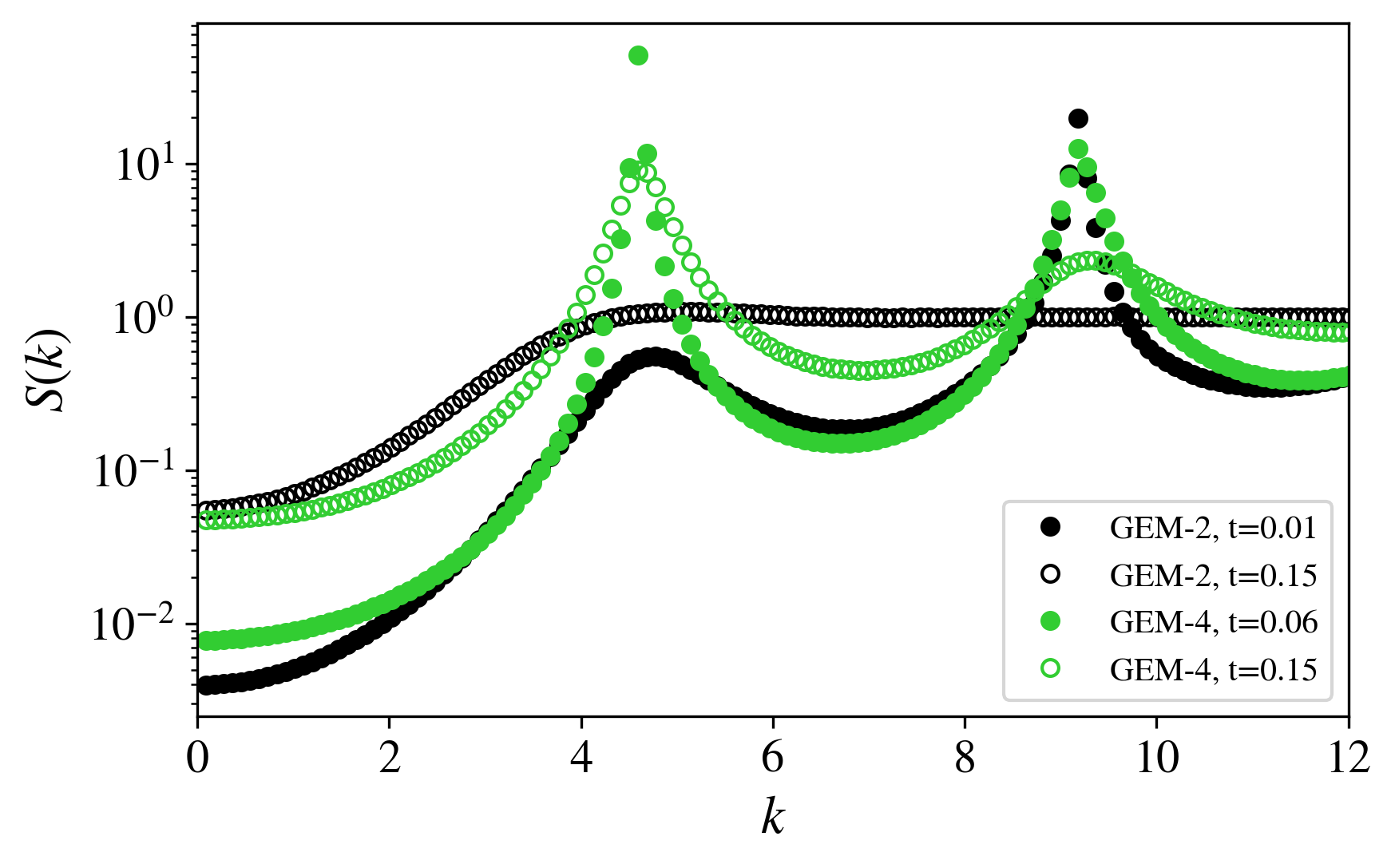}
  \caption{Comparison between the GEM--2 and GEM--4 static structure factors $S(k)$ at various temperatures, computed at reduced density 1.46165. Note the logarithmic scale on $S(k)$. Errorbars are smaller than the line width.}
  \label{fig:gem2vsgem4_sofq}
\end{figure}

Analogous physical information about the soft--particle behavior can be extracted from the static structure factor $S(k)$, where $S(k)=\frac{1}{N}\langle \rho_k \rho_{-k} \rangle$ with $\rho_k =\sum_{l=1}^N e^{ikx_l}$, in Figs. \ref{fig:sofq} and \ref{fig:gem2vsgem4_sofq}. The wavevector of the first peak in the $S(k)$ (Fig.~\ref{fig:sofq}) is given by the value of $k_{min}$ for the SH--6 potential, thus underlining that the forming structure is closely related to the two--particle clusters. Moreover, when temperature decreases, the appearance of the second peak at $k=2\,k_{min}$ highlights the tendency to the crystalline ordering among the clustering particles. 

We also compare, in analogy with the observations reported for the pair distribution functions, the static structure factors for the GEM--2 and GEM--4 systems at the same temperatures.
The peak of $S(k)$ located at $k_{min}$ shown in the picture is again a clear sign of the tendency to clustering; the behavior of the GEM--4 system displayed here (green points in Fig.~\ref{fig:gem2vsgem4_sofq}) is then similar to the one suggested by the $S(k)$ of the SH--6 potential. 
Concerning the small-$k$ peak of $S(k)$, at low temperature the GEM--2 also presents some structure which can be interpreted as a signal of the aborted clustering. Both models show the formation of a peak at $k=2 k_{min}$ which represents the tendency to solidification of clustering pairs of particles for the GEM--4 and of particles for the GEM--2.

\begin{figure}[tbp]
  \includegraphics[width=\columnwidth]{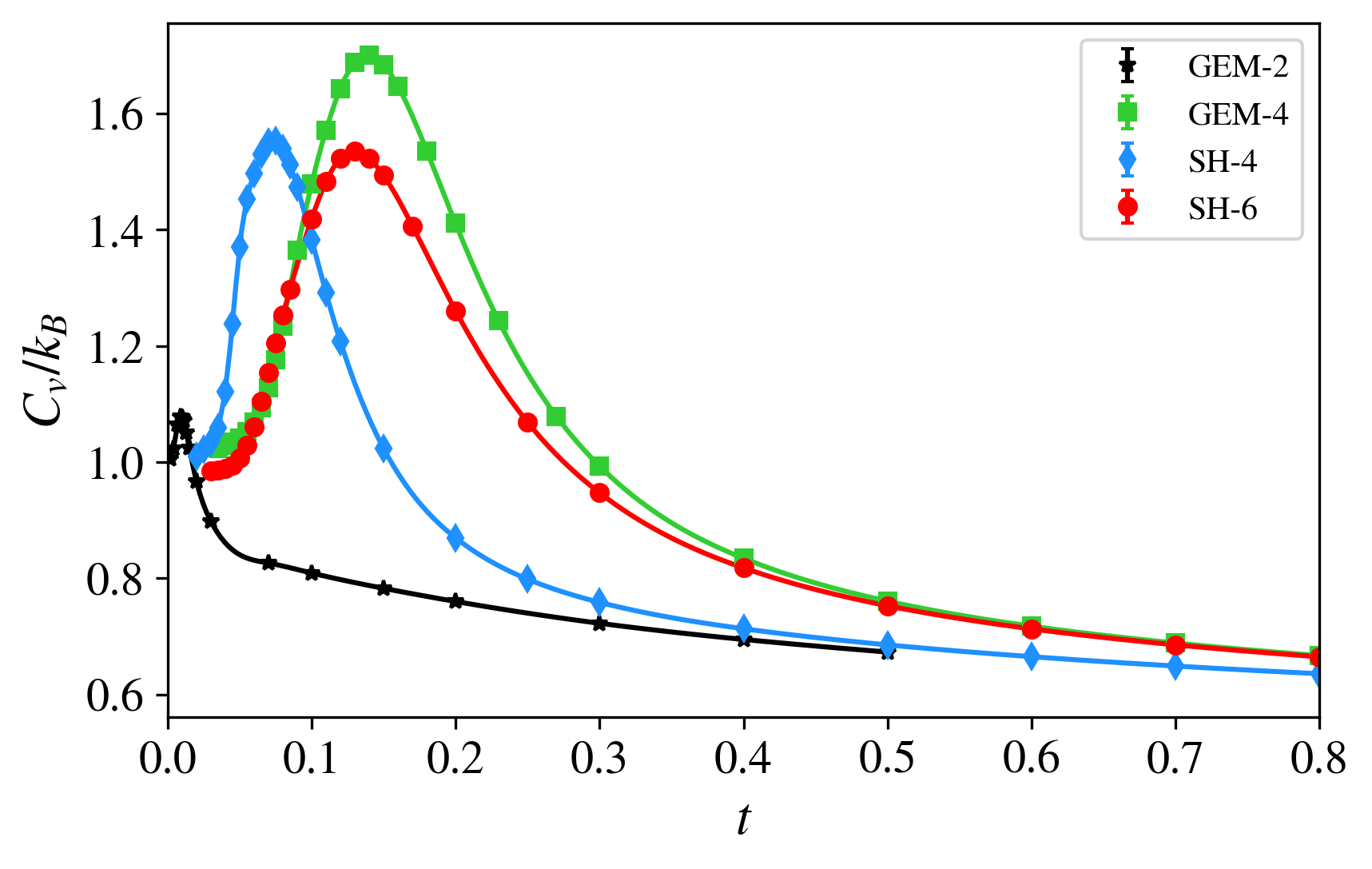}
  \caption{The specific heat at fixed volume, $C_v$, for GEM--4, SH--4 and SH--6 over a wide range of temperatures is reported. Also data of $C_v$ for the GEM--2 (non--clustering) potential are reported. Solid lines are guide for the eye.}
  \label{fig:cvsoft}
\end{figure}

We also calculate the specific heat for these systems. In particular, Fig.~\ref{fig:cvsoft} contains the simulated values of the specific heat $C_v = C/N$ in units of $k_B$, for the three aforementioned potentials, where $C$ is the heat capacity:
\begin{equation}
C=\dfrac{k_B(\langle V^2 \rangle- \langle V \rangle^2)}{t^2} + \frac{1}{2} \, N \, k_B\;.
\end{equation}
In this equation, $V$ is the potential energy of the soft-particle system and the second term is the analytical constant kinetic energy contribution to $C$, which is included for the sake of completeness.
The most striking feature in Fig.~\ref{fig:cvsoft} is that, at low temperatures, all the clustering potentials present a well--defined peak in the specific heat curves as a function of $t$. This behavior is akin to the Schottky anomaly of solid--state spin systems\cite{Gopal_Schottky}, which prompts the investigation of the emergence of spin degrees of freedom in the next Sections. The temperature range relative to the appearance of this peak is potential--dependent. The maximum heights of these curves have similar values for the three models investigated. On the contrary, the non--clustering GEM--2 system shows a much smaller peak at temperatures about one order of magnitude lower than the clustering potentials. 
A tendency to spatial ordering in the GEM--2 thus occurs at lower temperatures, via single-particle occupation of effective lattice sites. Conversely, the high and broad peaks present in the other three systems are a sign of the tendency towards the formation of two--particle clusters already at higher temperatures. This emergent phenomenon is more fragile in the SH--4 model, as shown by the $C_v$ peak, whose range is shifted towards lower temperatures. 
Note also that this peak, as well as the ones of the SH--6 and GEM--4 potentials, is much higher than that of the non--clustering potential, because of the larger energy fluctuations occurring in the clustering phenomena.
Approaching the limit of $t=0$, $C_v/k_B$ approximates the unit value, which corresponds to the value for an ideal harmonic solid; at high $t$, the simulated data display a convergence towards $1/2$, the typical value for the 1D ideal gas with only kinetic contribution.

\section{Mapping and pseudo--spin observables}\label{sec:mapping} 

Recently, the study of the SH--6 potential was tackled for a fluid of indistinguishable bosons in the continuum at zero temperature\cite{ref_sei,teruzzi2017,teruzzi2018,Prestipino2019}. The system was investigated upon changing the density and the interaction constant $U$. In the quantum regime, in fact, properties of the system do not depend only on the reduced temperature $t=k_B\,T/U$, but also on the dimensionless coupling constant $u=U m \sigma^2/\hbar^2$, where $m$ is the particle mass. The constant $u$ accounts for the relative role of interaction versus quantum delocalization effects. In particular, by fixing the density at the commensurate value $\rho=\rho_{(2)}=1.36857$, a QPT corresponding to the formation of two--particle clusters was detected at $u=u_c\simeq 18$\cite{ref_sei}. Below this value, a liquid regime was found, where ground state and excited state properties are typical of a Luttinger liquid of single particles, whereas for values above $u_c$ the behavior was the one of a two--boson cluster Luttinger liquid.
The transition between these two regimes, at $u=u_c$, is highlighted by a pair distribution function $g(x)$ which develops a peak at zero distance for $u>u_c$, but also by a marked anomaly in the Luttinger parameter, which characterizes the hydrodynamic properties of the fluid, across the transition. 
Interestingly, across the transition, an analysis of the excited states via extraction of the dynamical structure factor using analytic continuation\cite{gift1,gift2,gift3,gift4} demonstrates a gapped secondary mode connected to the incipient cluster formation. The gap of the secondary mode is found to vanish at the transition; this is the crucial clue for a further investigation of the possible mapping between this system and the quantum Ising model in transverse field, which has a similar behavior in the excited states' spectrum across its QPT\cite{pfeuty}. 

\begin{figure}[tbp]
  \includegraphics[width=\columnwidth]{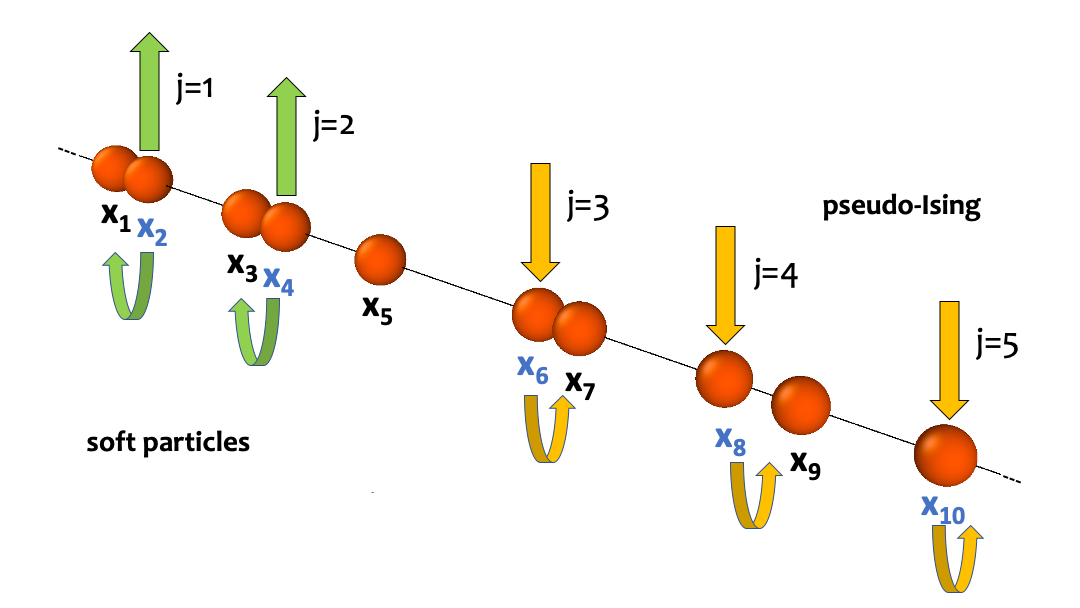}
  \caption{Schematic representation of the mapping between the soft--particle degrees of freedom and the corresponding pseudo--spins. The assignment of a left/right closest neighbor (lower curved arrows) is done for the particles with even indexes (labeled in blue), and determines the value of the pseudo spin (upper arrows). Considering the odd particles leads to an equivalent result.}
  \label{fig:mapping}
\end{figure}

This mapping can be made quantitative by introducing a set of string variables similar to the one discussed in Ref.~\cite{Ruhman2012}: first, particles are ordered by their increasing position, $x_1 < x_2 < \dots < x_{N}$ and even positions are assigned a lattice index $j=1, \dots, N/2$. Second, a pseudo--spin $\sigma_{j}=1$ is assigned if $\left|x_{k}-x_{k-1}\right|<\left|x_{k}-x_{k+1}\right|,$ or $\sigma_{j}=-1$ in the opposite case, with $k=2j$; note that when these operations involve particle 1 and particle $N$, these are first neighbors in PBC.
The last step of the mapping consists in placing these spins onto evenly spaced lattice sites, thus neglecting the fluctuations of the particle coordinates in the original soft system. 
A fully equivalent mapping can be done by using only the odd-indexed particles as starting points.
The whole procedure is sketched in Fig.~\ref{fig:mapping}.
As a consequence, it is possible to directly associate these $N_{\text{s}} = N/2$ discrete spin variables to the $N$ soft continuous configurational degrees of freedom. This mapping is a surjective function of the spatial coordinates and performs a coarse graining of the configurations, leaving only the relevant "magnetic" degrees of freedom. Note that a completely disordered configuration of soft particles, being the distance between a particle and its two neighbors a random value, would be mapped onto a randomly--oriented pseudo--spin configuration. Conversely, an almost perfectly two--particle clustered configuration would be mapped onto a spin configuration with all the spins aligned in the same direction. The order parameter is essentially the thermal average of these pseudo--spins.
Moreover, as it is exemplified in Fig.~\ref{fig:mapping}, it could happen that a pseudo--spin with a given orientation is followed by a pseudo--spin with the opposite orientation. 
Remarkably, in the 1D quantum soft system, the pseudo--spin correlation function $\langle\sigma_{j}^{z} \sigma_{j+i}^{z}\rangle$ (averaged over $j$) behaves as expected for the transverse Ising model. In the paramagnetic phase, which corresponds to a Luttinger liquid of soft particles, it decays exponentially; conversely, in the ferromagnetic phase, which corresponds a two--particle cluster Luttinger liquid, it manifests true long-range order. This procedure allows then to observe the order emerged in this QPT, which is a nonlocal form of ordering evidenced by the procedure employed for mapping the soft system onto a system of pseudo--spins hosted on a regular 1D lattice.
All the positions of the soft particles are involved, indeed, in the construction of the string variables: the particles must, in fact, be labelled in increasing order, to be able to define each pseudo--spin via the mapping algorithm.

\begin{figure}[tbp]
  \includegraphics[width=\columnwidth]{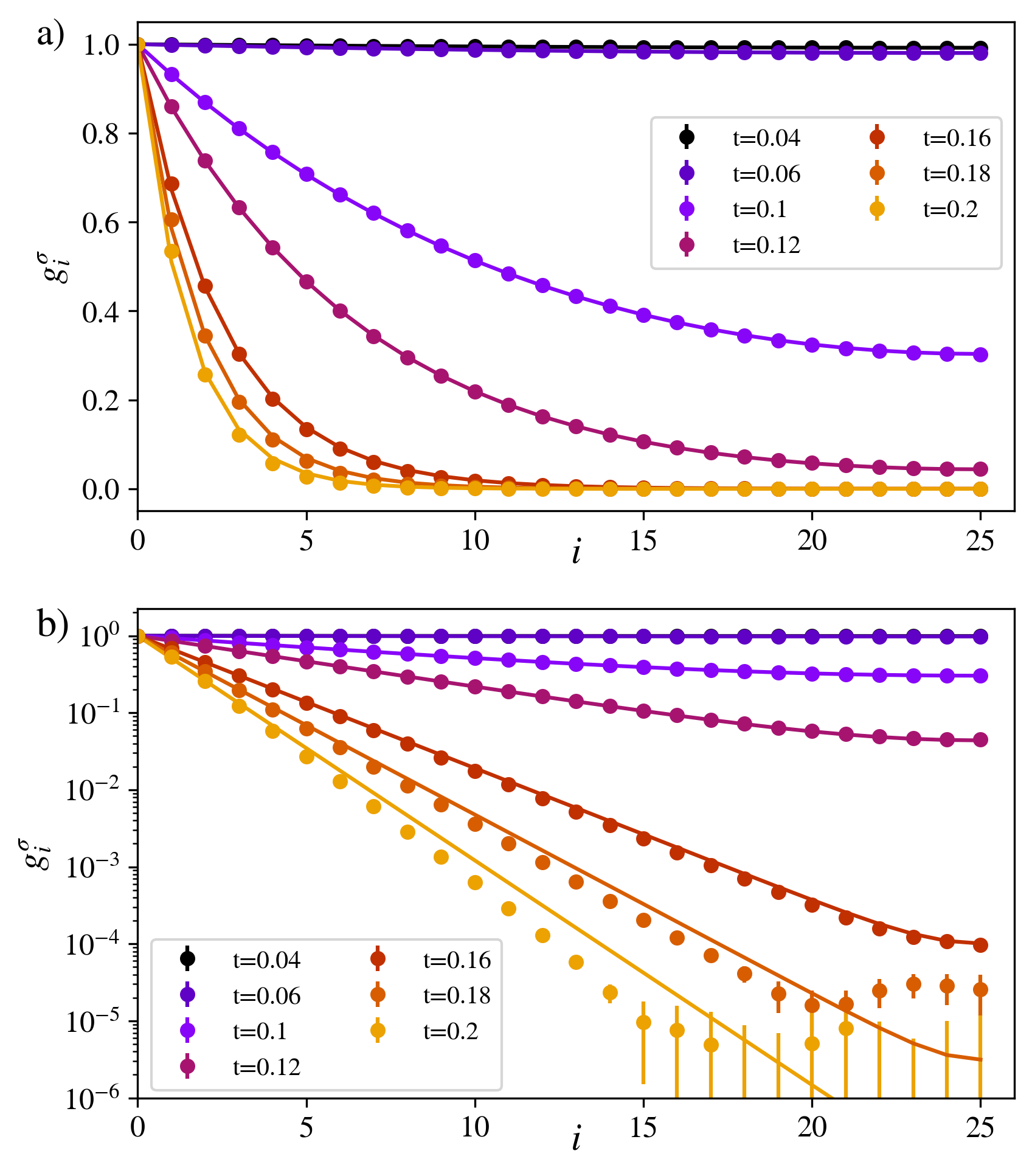}
  \caption{a): Ising correlation functions for pseudo--spins $g^\sigma$ are shown in the case of the GEM--4 interaction. Solid lines represent, at each temperature, independent fits with the exact Ising formula up to half the size of the effective lattice. b): same data in linear--logarithmic scale with errorbars.}
  \label{fig:correl}
\end{figure}

Concerning the classical 1D soft systems under investigation in this work, a phase transition at finite temperature is not expected, as discussed above; however, at very low temperatures, our systems enter a critical regime corresponding to a tendency to clustering which becomes effective only for $t=0$. Moreover, the appearance of a Schottky-like anomaly in the specific heat suggests that a discrete number of well-defined metastable states contributes in the low-temperature dynamics of clustering systems. An intriguing possibility is that these soft systems could as well be partially mapped onto the 1D Ising model, but in this case a classical one.  This would imply that the part of the dynamics of a generic soft 1D system on the continuum, pertaining to clustering effects, could be mapped onto a system of discrete variables on a lattice. We notice that there are many instances in statistical physics where a continuos system has the same critical behavior as a discrete model, for example the three-dimensional (3D) liquid-gas transition is in the 3D Ising class. Moreover, discrete variables are customarily introduced in the study of clustering systems, describing the number of particles per cluster. Here, the role of magnetic discrete degrees of freedom can be directly tested by using the mapping of the soft--particle degrees of freedom onto the pseudo--spin variables already exploited in the quantum case.
To this aim, via the mapping procedure just introduced, it is possible to calculate the Ising--like thermodynamic properties of the pseudo--spins. A major advantage of this approach is that it is well-defined for all potentials and temperature regimes that we are considering, without any approximation.
We are interested not so much in the characterization of the thermodynamic limit, as in the comparison between the pseudo--spin observables computed in the simulations and the analytical calculations for an analogous Ising system with the same number of spins. 

\begin{figure}[tbp]
  \includegraphics[width=\columnwidth]{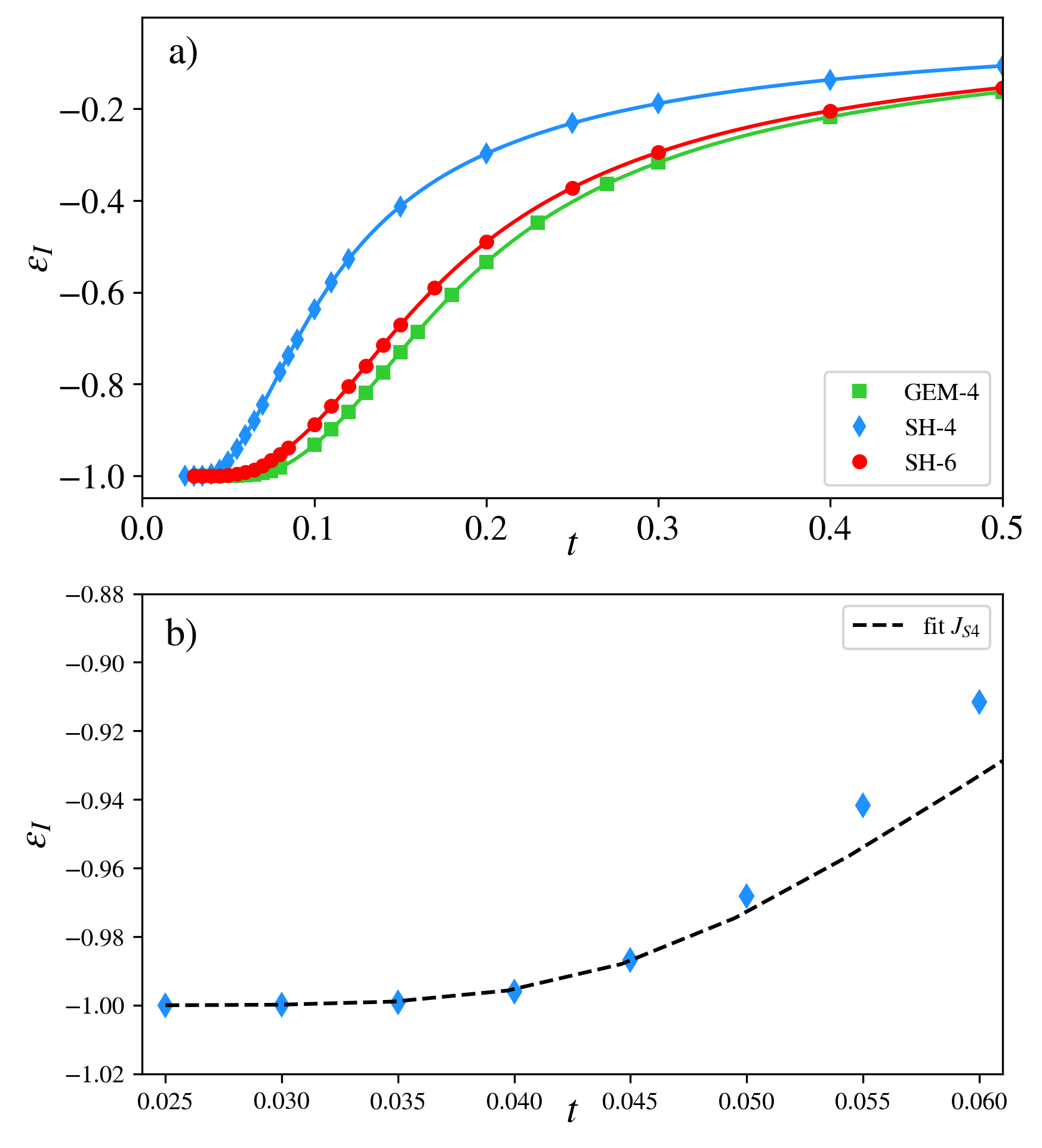}
  \caption{Ising energy per spin in units of $J$, obtained from soft--particle configurations via mapping. Solid lines are guide for the eye. The lower panel shows a closer view of the temperature range between 0.025 and 0.06, only for SH--4, and the black dashed line is the Ising result with $J=J_{S4}$.}
  \label{fig:ising_energy}
\end{figure}

In parallel to the approach adopted in the study of the 1D quantum system, the first step consists in applying the mapping procedure for the computation of the pseudo--spin correlation functions averaged over the starting index, $g^\sigma_i=\sum_j\langle \sigma_j \sigma_{j+i} \rangle/N_{\text{s}}$. As an example, in Fig.~\ref{fig:correl} we report a relevant subset of those computed for the GEM--4. The following considerations hold however for all the three interaction potentials studied in this work. Remarkably, the shape of all these curves closely resembles the behavior expected for a classical Ising model with short range interactions, but with a coupling constant $J$ dependent on temperature. Note, in fact, that we necessarily have to associate the temperature $t$ of the soft system to the one of the pseudo--spin model. This implies that we let the Ising coupling $J$ (assumed to be in $U$ units) as a free parameter to be fitted at each temperature.   
Clearly, as evidenced by the lin--log scale of Fig. \ref{fig:correl}b) the fit turns out to be very accurate at low $t$; in particular, at $t=0.16$ and below the simulated data are fitted quite closely by the theoretical curve, while at higher temperatures the mapping procedure does not generate pseudo--spin variables effectively associable to Ising spins. In fact, the simulated data points deviate from the model, at large $i$ values, for $t=0.18$ and $t=0.20$. Noticeably, as it will be shown in the following (see Fig. \ref{fig:jeffofTg4}), a further temperature lowering is required for pseudo--spin behavior to closely manifest Ising physical properties. As in the Ising model, the pseudo--spins appear to be strongly correlated at very low temperatures, where $g^\sigma$ is almost flat and slowly decaying; conversely, the pseudo--spins start to assume random relative orientation, as the temperature grows and $g^\sigma$ rapidly decreases towards zero. 

\begin{figure}[tbp]
  \includegraphics[width=\columnwidth]{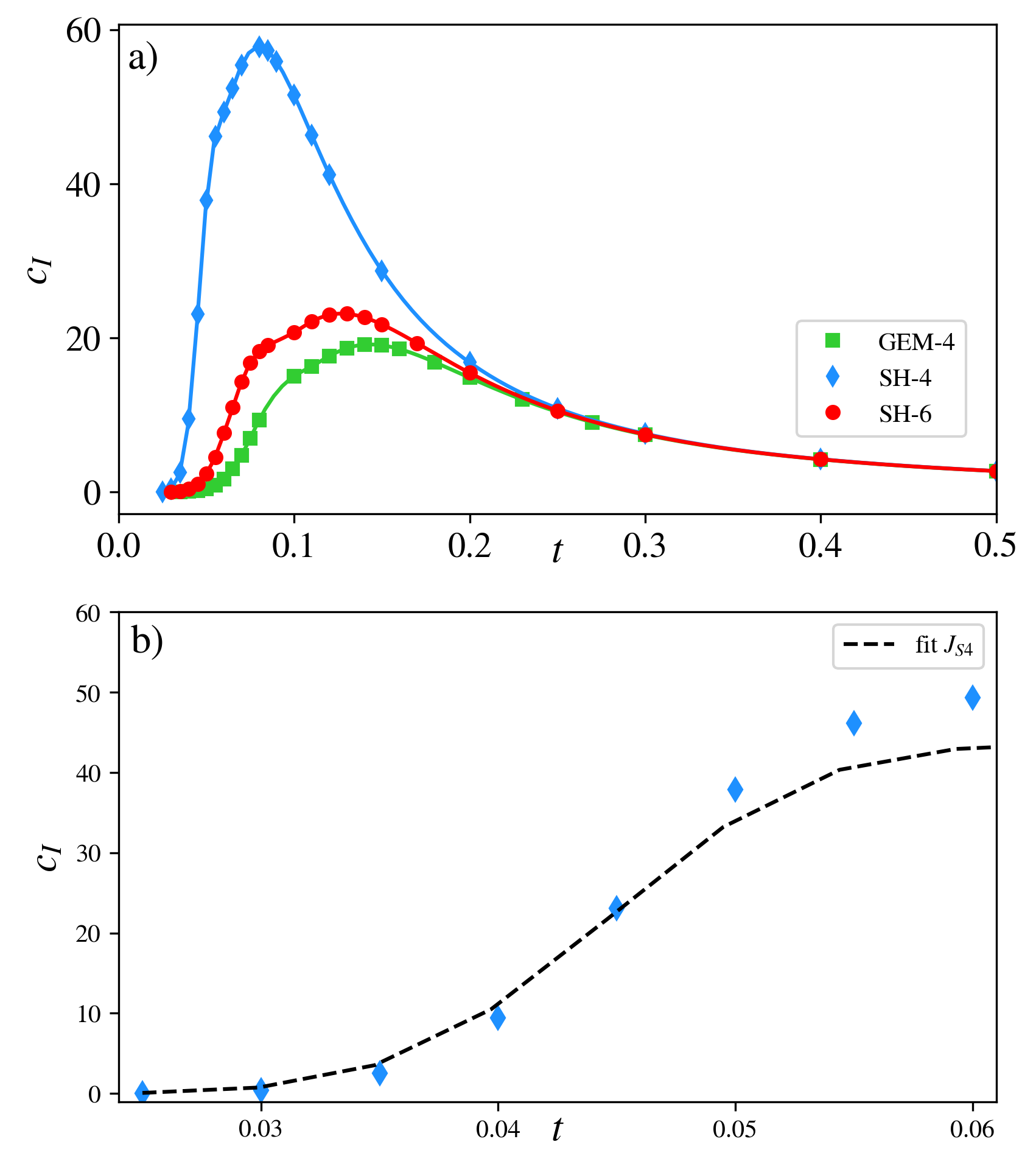}
  \caption{Ising specific heat in units of $k_B J^2$, obtained via mapping from soft--particle configurations. Solid lines are guide for the eye. The lower panel shows a closer view of the temperature range between 0.025 and 0.06, only for SH--4, and the black dashed line is the Ising result with $J=J_{S4}$.}
  \label{fig:ising_specific_heat}
\end{figure}

\begin{figure}[btp]
  \includegraphics[width=\columnwidth]{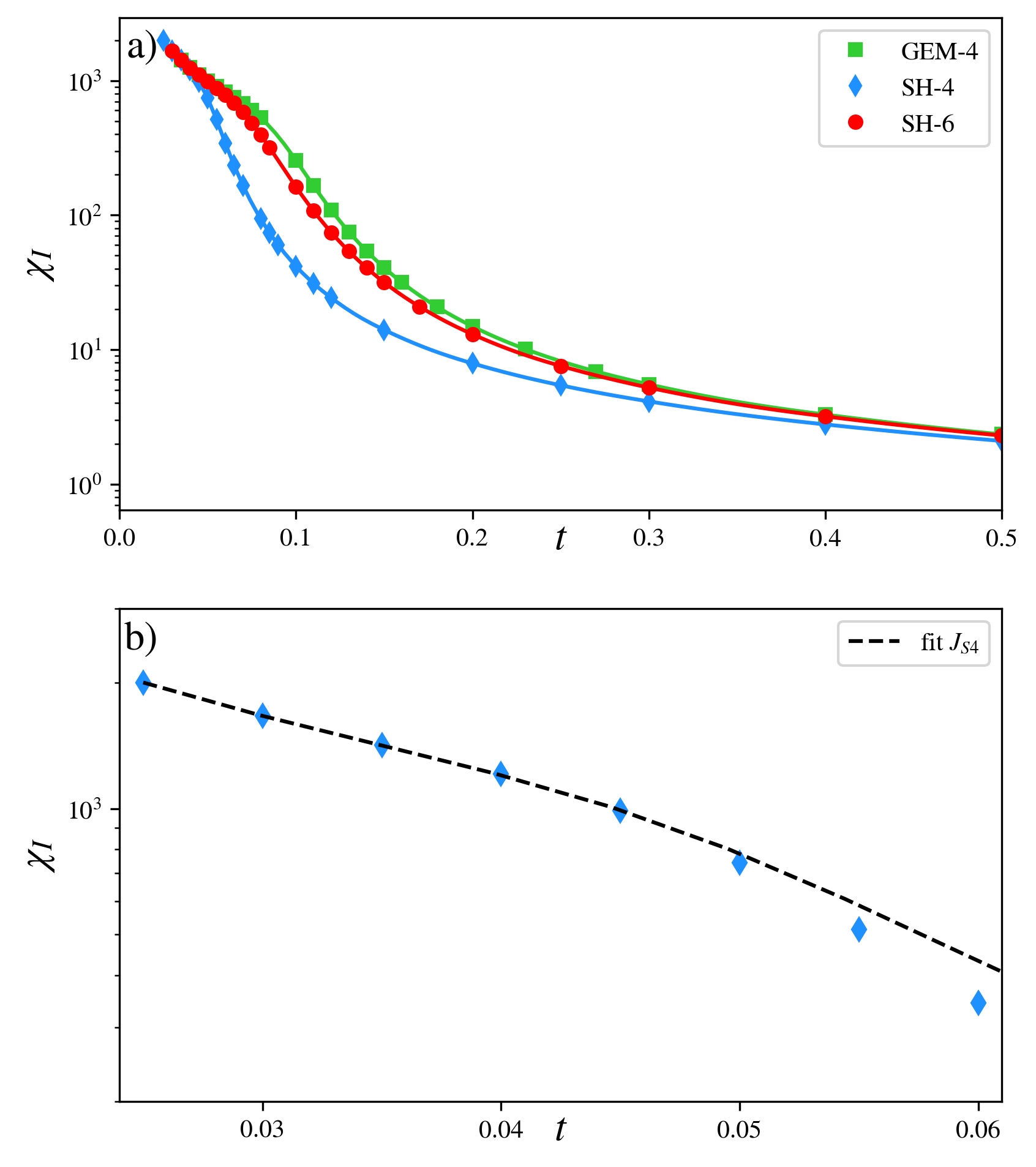}
  \caption{Ising magnetic susceptibility, obtained via mapping from soft--particle configurations. Solid lines are guide for the eye. The lower panel shows a closer view of the temperature range between 0.025 and 0.06, only for SH--4, and the black dashed line is the Ising result with $J=J_{S4}$.}
  \label{fig:ising_susceptivity}
\end{figure}

The study of the $g^\sigma$ functions suggests that the hypothesis of a mapping between the soft system and the Ising model can be quantitative. As a further step in this direction, we also compute other relevant thermodynamical observables of the mapped pseudo--spin system. In Figs. \ref{fig:ising_energy}, \ref{fig:ising_specific_heat} and \ref{fig:ising_susceptivity}, we show, for all the three interaction models, the pseudo--spin properties of energy, heat capacity and susceptibility, obtained by mapping the configurations of $N=100$ soft particles onto ${N_{\text{s}}}=50$ pseudo--spins. 
We define the Ising total energy as $E_{I}=-J\,\sum_{i=1}^{N_{\text{s}}} \sigma_{i} \sigma_{i+1}$ (with PBC) and the magnetization as $M=\sum_{i=1}^{N_{\text{s}}} \sigma_{i}$. Since $J$ is an unknown parameter, in our simulations we evaluate the pseudo--spin thermodynamic quantities by suitably factoring out the trivial $J$ dependence, and thus using the following formulae: 
the energy per spin, in units of $J$, $\varepsilon_I=E_{I}/N_{\text{s}}J$; the specific heat in $k_B \, J^2$ units
$c_I=N_{\text{s}} \left(\langle \varepsilon_{I}^{2} \rangle - \langle \varepsilon_{I} \rangle ^{2}\right)/ t^2$; and the magnetic susceptibility, $\chi_I= \langle M^{2} \rangle/N_{\text{s}}\,t$, where we use $\langle M \rangle=0$ since we are in the paramagnetic phase.

As an example of the critical Ising regime unveiled via mapping with the string variables, in the insets of Figs. \ref{fig:ising_energy}, \ref{fig:ising_specific_heat} and \ref{fig:ising_susceptivity} we also show a curve (black dashed line) obtained by simultaneously fitting $\varepsilon_I, c_I$ and $\chi_I$ for the SH--4 potential to the corresponding theoretical expressions for the 1D Ising model with nearest--neighbor interaction and PBC (recalled in the Appendix). The fit is done in the $t$-range 0.025-0.06 and yields a value $J=J_{S4}=0.099235$. The results highlight a fairly good agreement between the quantities obtained via mapping and the analytical expressions, by using the same $J_{S4}$ for all the observables. However, this only holds at very low temperatures; upon increasing temperature, in fact, the data points start to deviate from the fitted curve. This is indeed what one should expect, given that the critical regime (marked by the peak in che specific heat) is approached for $t\to0$. The simple first--neighbor Ising model should not be able to adequately fit a whole range of finite temperatures with a single parameter ($J_{S4}$, in this case). Therefore, we could expect that the mapping onto the Ising model only becomes exact in the limit $t \rightarrow 0$ and that an effective mapping at finite temperature has the spin--spin coupling constants (even beyond first neighbors) significantly dependent on temperature. 

The Ising--like energy of the pseudo--spins system reaches -1 in the $t \rightarrow 0$ limit, while it increases for higher temperatures, as expected. Also $c_I$ and $\chi_I$ behave like the corresponding observables of a Ising spin system with first--neighbor interactions, but the quantitative agreement with this model only holds at very low temperatures (see Figs. \ref{fig:ising_energy}b), \ref{fig:ising_specific_heat}b) and \ref{fig:ising_susceptivity}b)), thus highlighting that the soft system at high temperatures departs from the critical regime. The specific heat tends to 0 as $t \rightarrow 0$, and then it grows up to a maximum value located between $t=0.1$ and $t=0.15$, depending on the soft interaction considered; then, it starts to decrease. For $t \rightarrow 0$, the susceptibility experiences a considerable growth (note the $y$--log scale), revealing a magnetic ordering of the pseudo--spin system. 

\section{Search for the Hamiltonian coupling constant}\label{sec:jeff} 

\begin{figure}[tbp]
  \includegraphics[width=\columnwidth]{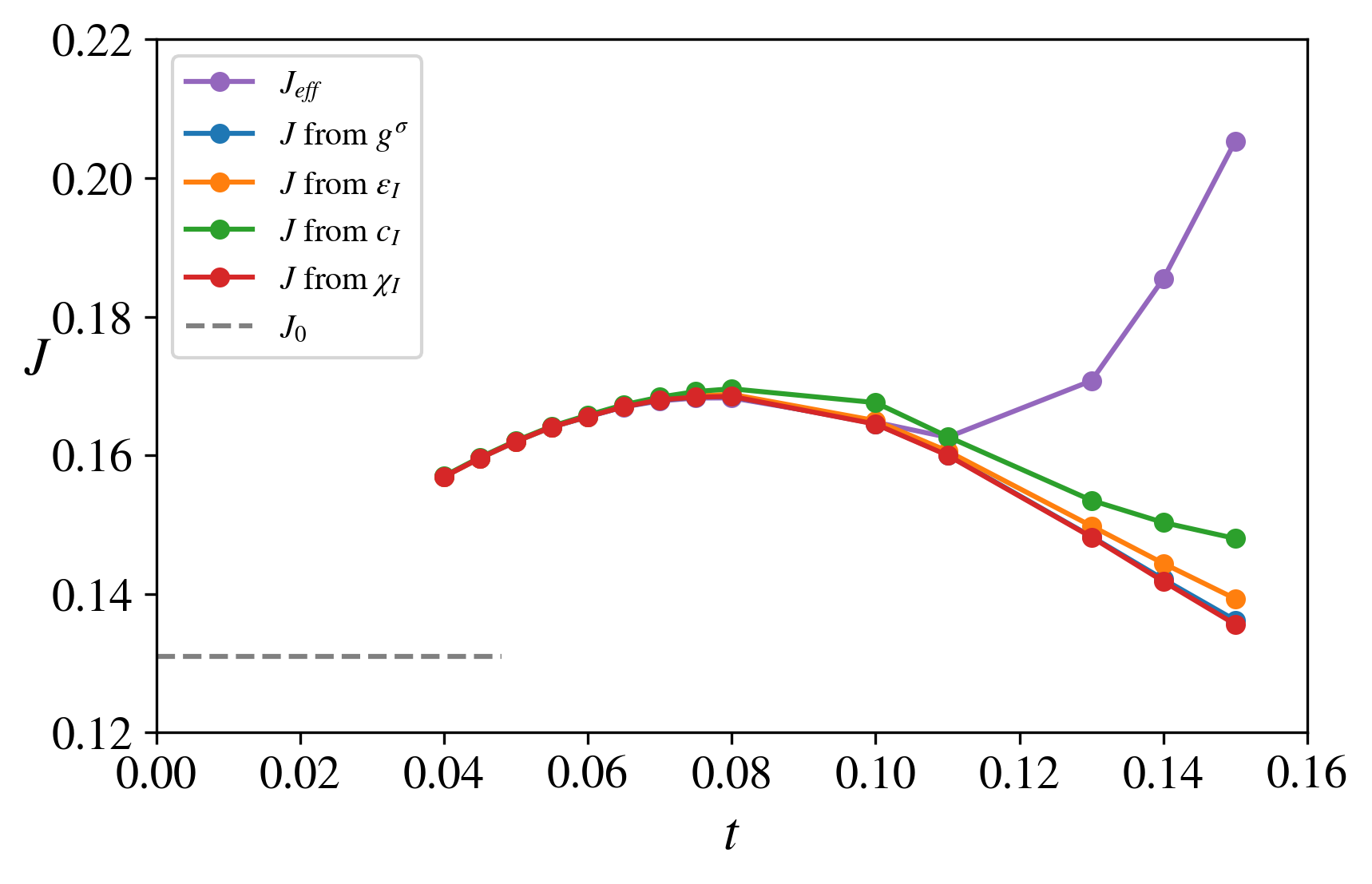}
  \caption{For the GEM--4 potential, estimates of coupling constant $J$ as a function of $t$ extracted from the pseudo--spin thermodynamic properties, from the defects' occurrence analysis and from SA. Solid lines are guide for the eye.}
  \label{fig:jeffofTg4}
\end{figure}

\begin{figure}[tbp]
  \includegraphics[width=\columnwidth]{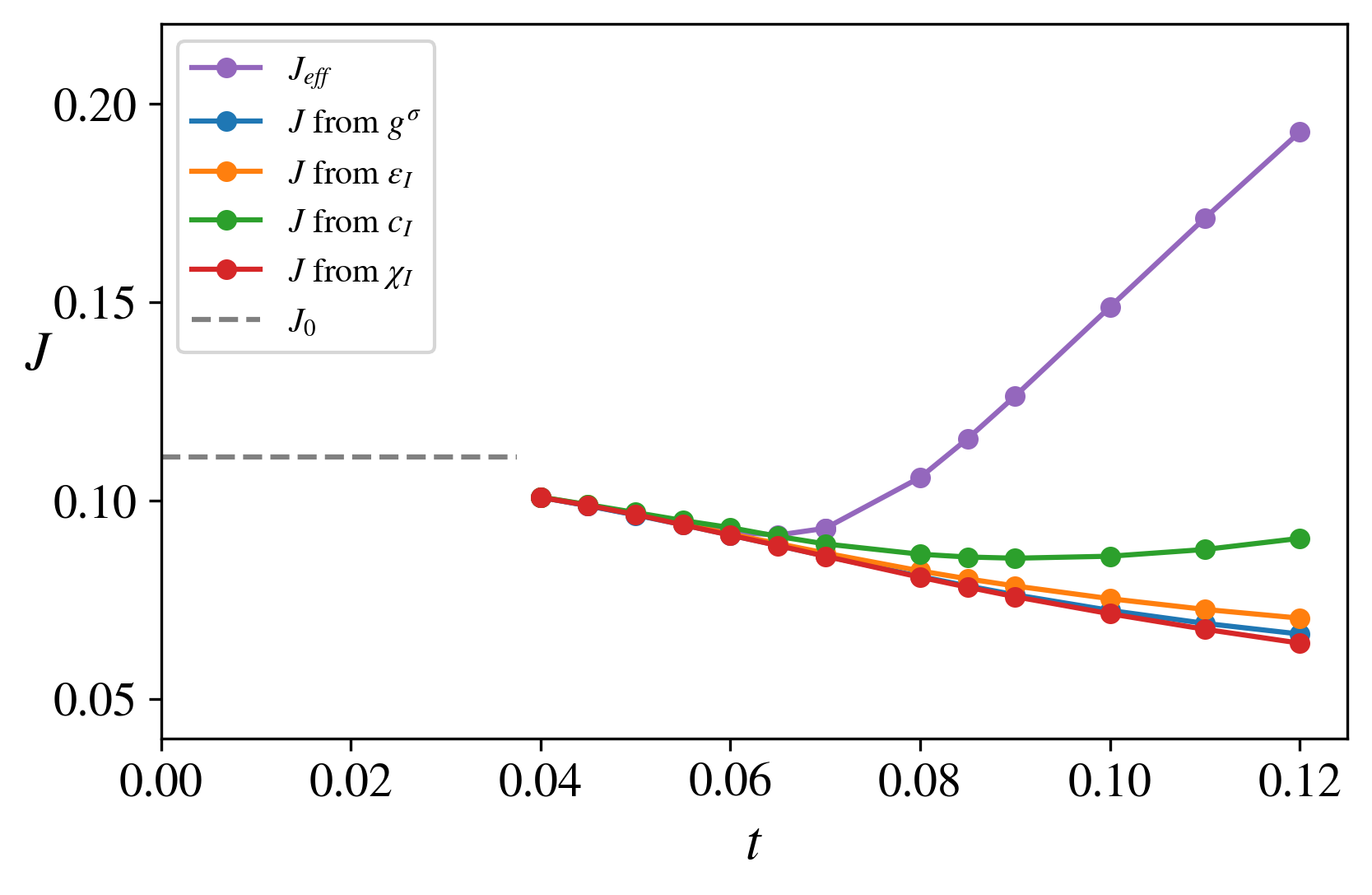}
  \caption{Estimates of $J$ for potential SH--4, analog to Fig.~\ref{fig:jeffofTg4}.}
  \label{fig:jeffofTs4}
\end{figure}

\begin{figure}[tbp]
  \includegraphics[width=\columnwidth]{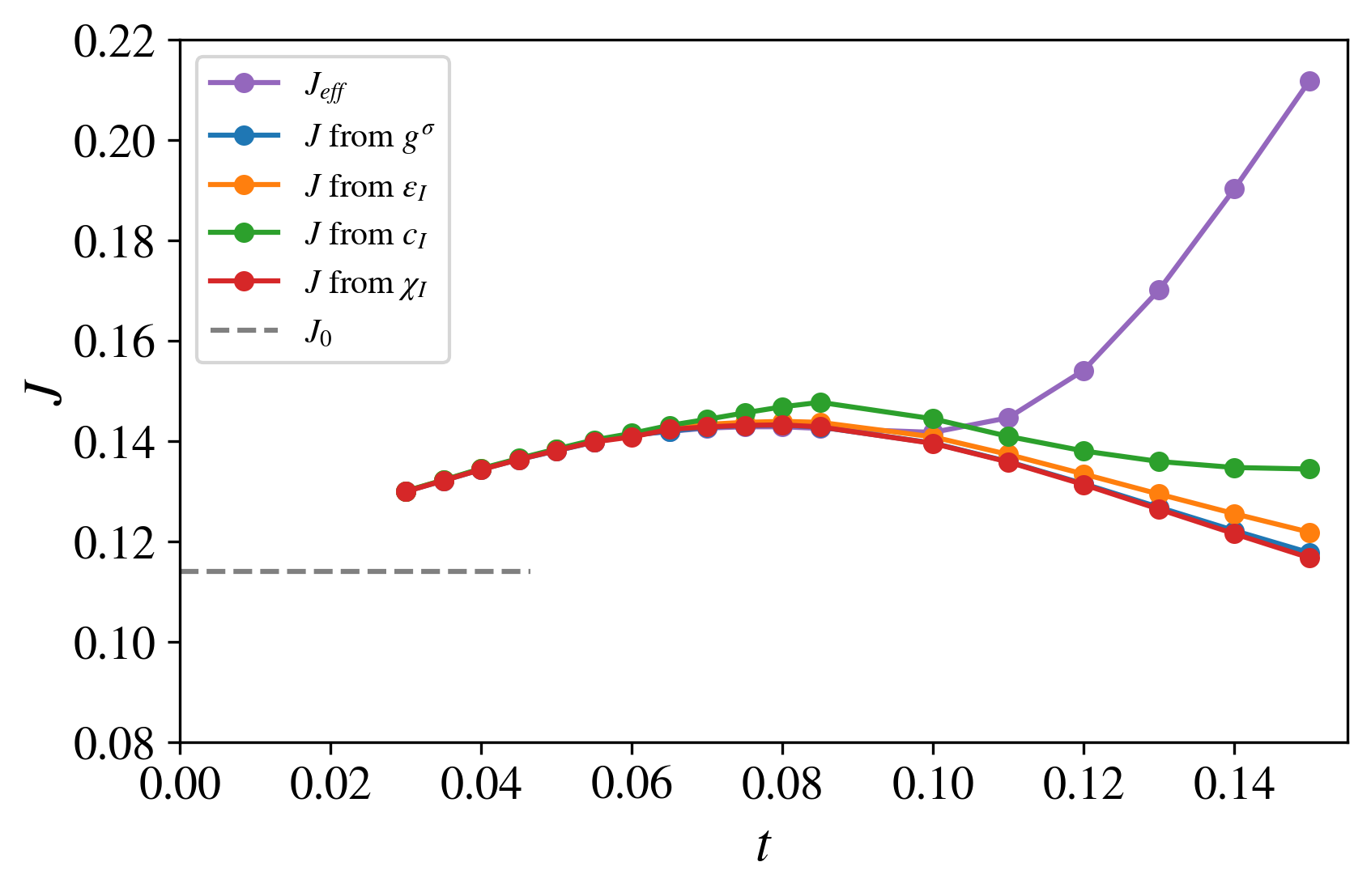}
  \caption{Estimates of $J$ for potential SH--6, analog to Fig.~\ref{fig:jeffofTg4}.} 
  \label{fig:jeffofTs6}
\end{figure}

We have shown that the critical low--temperature regime (identified by the specific heat behavior) of the soft systems under investigation can be approximately mapped onto an Ising model with first--neighbor interactions via a temperature dependent coupling constant. This procedure effectively traces out the continuous phononic degrees of freedom, and the result is very interesting, because it shows that also in the 1D classical case, on approaching the zero--temperature clustering regime, a mapping onto a discrete Ising model can be found, starting from a continuous fluid. 
It is therefore possible to compare the data in Figs. \ref{fig:correl}-\ref{fig:ising_susceptivity} to the theoretical curves for the Ising model with $N_{\text{s}} = 50$.
This way, the value of $J$ as a function of $t$ can be determined for each observable as follows. Concerning specific heat, energy and susceptibility, we determine the value of $J$ for which the theoretical curve exactly coincides with the simulated data at each $t$. On the other hand, pseudo--spin correlation functions directly provide a $J$ value for each $t$ via the fit with the exact Ising formula for $g^\sigma$, as detailed in the previous Section. This procedure provides a set of discrete data points which are drawn in Figs.~\ref{fig:jeffofTg4}, \ref{fig:jeffofTs4} and \ref{fig:jeffofTs6} for the three interaction models. At high temperatures, the estimates of $J$ extracted from different physical observables are different from each other. Quite remarkably, by lowering $t$, the four curves collapse onto a unique curve that still displays a dependence on temperature. In the light of the exponential decay of the spin--spin correlation function, it is not surprising that it can be fitted by its Ising form with a suitable value of $J$. However, it is noticeable that this precise value is almost identical to the three $J$ values estimated from the specific heat, energy and susceptibility.
This holds for all the potentials studied and this remarkable collapse marks the entry in the critical regime.

\begin{figure}[tbp]
  \includegraphics[width=\columnwidth]{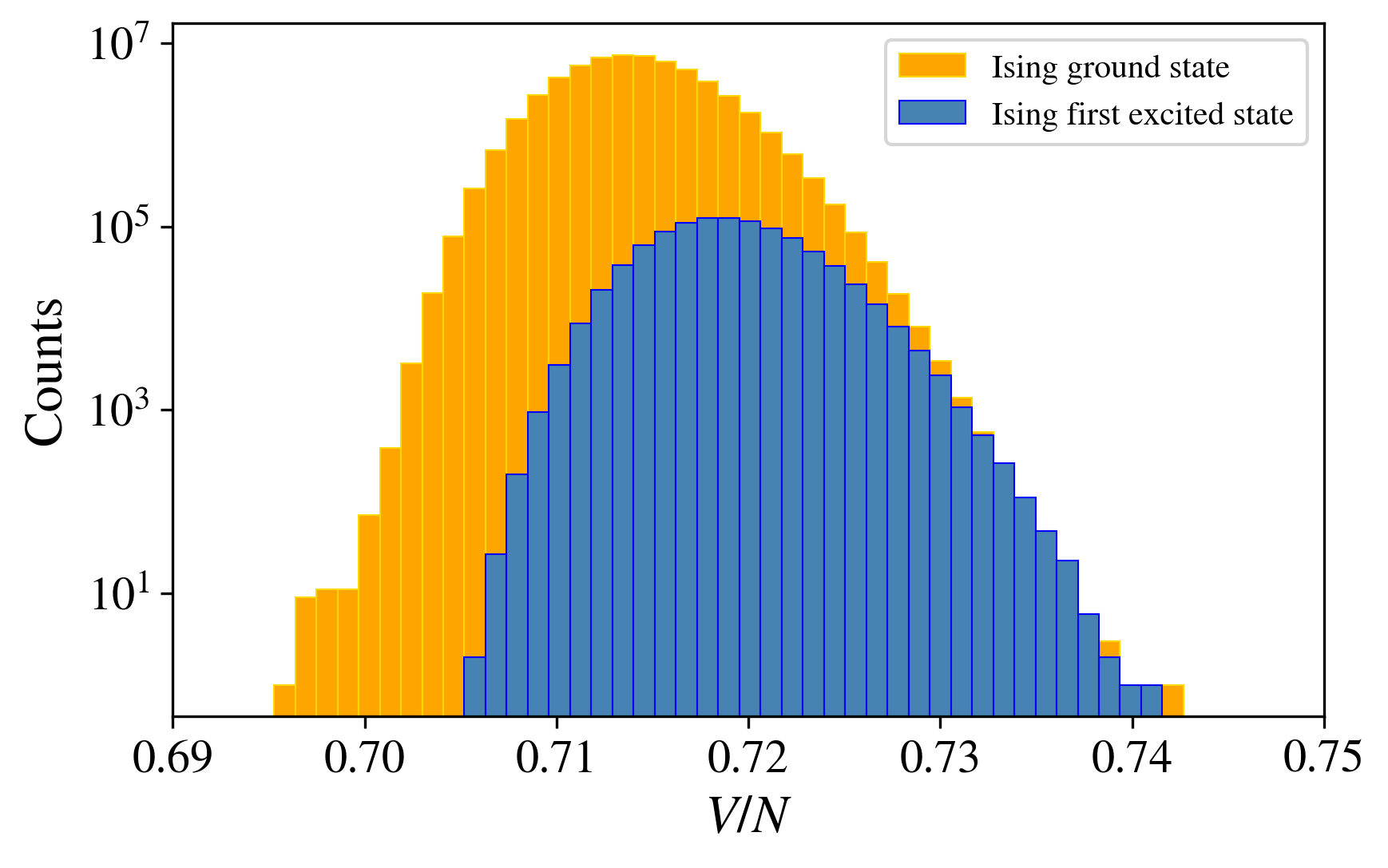}
  \caption{Histogram of the soft potential energy per particle, $V/N$ for SH--6 at $t=0.05$. The energies of the soft configurations belong either to the group of those mapped onto a Ising ground state (yellow) or to the group of those with two domain walls in the corresponding Ising system (blue).}
  \label{fig:histogram}
\end{figure}

At this point, it is fundamental to assess whether the mapping of the soft system onto the Ising model is trivial, i.e., if only low--energy soft configurations are sent onto the pseudo--spin ground state, while only high--energy soft configurations are mapped onto the Ising states with defects.
Fig.~\ref{fig:histogram} shows, as an example, two histograms of the soft particles energies from a $t=0.05$ simulation of the SH--6 interaction model. These distributions collect all the potential energy values of the soft configurations sent onto the Ising ground state (yellow) or onto an Ising first excited state (blue), i.e. a pseudo--spin configuration with two domain walls. In this analysis, the soft configurations mapped onto more defected Ising states have not been considered.
Evidently, there is a substantial overlap between the yellow and the blue histograms and not two neatly distinguishable distributions, which means that the mapping is highly nontrivial. This strongly indicates that the mapping procedure via the string variables establishes a complex relation between the physical system of soft particles and the equivalent system of pseudo--spins. We however expect that such distributions become less and less overlapping as temperature decreases to regimes which are not accessible to our simulations.

Thus, the characterization of the $t \rightarrow 0$ limit of the mapping, and then of the coupling constant $J$, becomes an interesting further step. At such low temperatures, the kinetic energy contribution to the Hamiltonian is negligible and the potential energy term dominates. In these conditions, an energy variation in the soft system is directly associable to an energy variation in the pseudo--spin mapped system. This provides us with an indirect method for the determination of $J$.
In fact, let us recall that the energy cost for a single domain wall in the Ising model with nearest--neighbor coupling is equal to $2J$. Considering PBC, the first excited Ising state has two domain walls in our simulations. This means that the minimum cost of a defect in the pseudo--spin model is equal to $4J$. The integral of the occurrence distribution of soft configurations that map to this lowest energy defected pseudo--spin configurations, at a given temperature, can thus be matched to a Boltzmann weight with potential energy equal to exactly $4J$.
This means that the relative probability of the first exited Ising state with respect to the Ising ground state, written as $e^{-4J/t}w_{\text{def}}/w_{\text{gs}} $, can be identified as the ratio between the number of fluid configurations that are sent into the first state and into the Ising ground state. In particular, this last fraction reads as $\langle e^{-(V_{\text{def}}-V_{\text{gs}})/t} \rangle$, where the energies $V_{\text{gs}}$ and $V_{\text{def}}$ are the total potential energies (i.e. $\sum_{i<j}^N v(|x_i-x_j|)$) of the soft configurations which are mapped, respectively, onto the Ising ground state and onto the first excited Ising state, i.e. the one with two domain walls, while $w_{\text{def}}=N_s (N_s-1)$ and $w_{\text{gs}}=2$ are the degeneracies of, respectively, the defected and ground states in the 1D Ising model with PBC. The right member can be determined by assuming that - for a long MC simulation - $\langle e^{-V_{\text{gs}}/t} \rangle$ is proportional to the total number of configurations sent onto the Ising ground state, $n_{\text{gs}}$, namely the sum of an histogram like the one in Fig.~\ref{fig:histogram}, and the same holds for the lowest--energy defected state (with the corresponding $n_{\text{def}}$).
The previous considerations allow us to introduce the definition of an \textit{effective coupling constant} $J_{\text{eff}}$ as:
\begin{equation}
J_{\text{eff}} = \frac{t}{4} \ln \Bigl(\dfrac{n_{\text{gs}}}{n_{\text{def}}} \, \dfrac{w_{\text{def}}}{w_{\text{gs}}} \Bigr)
\label{eqn:jeff}
\end{equation}
where $J_{\text{eff}}$ depends on $t$.
Therefore, this method not only provides the numerical value of $J$ in the $t \rightarrow 0$ limit, but also gives a physical insight into the coupling constant relation with the soft--particle properties. 
$J_{\text{eff}}$ as a function of temperature in the range of interest is shown in purple color in Figs.~\ref{fig:jeffofTg4}-\ref{fig:jeffofTs6}. These curves differ from the ones computed from the soft system observables at high $t$, but noticeably, at low temperatures, they collapse onto the other ones, despite being computed in a completely different way. Remarkably, the two different approaches yield almost indistinguishable results for the estimation of $J(t)$. This establishes a link between the thermodynamic properties of the pseudo spins and those of the soft particles.

We have also extracted the $t \rightarrow 0$ limit value for $J$ using SA\cite{kirkpatrick:1983}. 
The SA method allows to identify the soft--particle configurations corresponding to the lowest energy excitation of the system, which are mapped onto an Ising state with a single defect (namely two domain walls, in PBC). Starting at low temperature from a non perfectly clustered soft--particle configuration, the system is annealed towards the $t = 0$ limit, storing the positions of the particles and their potential energy. From this analysis, we have obtained that the lowest energy defect corresponds to an arrangement of two single particles and ($N_{\text{s}}-1$) two--particle clusters. 
We observe that the defect is highly localized: the particles immediately near the two single particles experience a huge displacement with respect to their position in a perfect lattice, due to the local depletion of the density. In turn, a few lattice steps away from the defect, the displacement are very small, and consistent with a slighlty higher uniform cluster density. Where the single particles are found, the relative pseudo--spin is flipped; the pseudo--spin is reverted again when the other defect is encountered. Consistently, starting from the first single particle, one of the two particles of each cluster is moved into the next cluster. The energy of the defect as a function of the distance between the two single particles rapidly tends to a constant value which characterizes the energy gap with respect to the ground state. The defect energy thus corresponds to a well--defined local minimum of the potential energy landscape, which is relevant at low temperatures since it provides metastability.
With this energy gap, we computed a temperature--independent $J=J_0$ value via the relation $J_{0} = (V_{\text{def}} - V_{\text{gs}})/{4}$, for each model. These $t \rightarrow 0$ limit values of $J$ are included in Figs.~\ref{fig:jeffofTg4}-\ref{fig:jeffofTs6} as horizontal gray dashed lines, and are also listed in Tab.~\ref{tab:simann}, for $N_{\text{s}}=50$. Ideally, all the $J$ values estimated from the soft physical observables and $J_{\text{eff}}$ should approximately tend, at low temperatures, to $J_0$. Reasonable extrapolations for $t \rightarrow 0$ of the previously discussed $J$ and $J_{\text{eff}}$ curves seem to be in good agreement with this hypothesis.

\begin{table}[tbp]
\begin{ruledtabular}
\begin{tabular}{ll}
Pair potential & $J_0$\\
\hline
GEM--4 & 0.130975\\
SH--4 & 0.110975\\
SH--6 & 0.114025\\
\end{tabular}
\end{ruledtabular}
	\caption{$J_0$ values calculated via Simulated Annealing for $N_{\text{s}}=50$.}
	\label{tab:simann}
\end{table}

\begin{figure}[tbp]
  \includegraphics[width=\columnwidth]{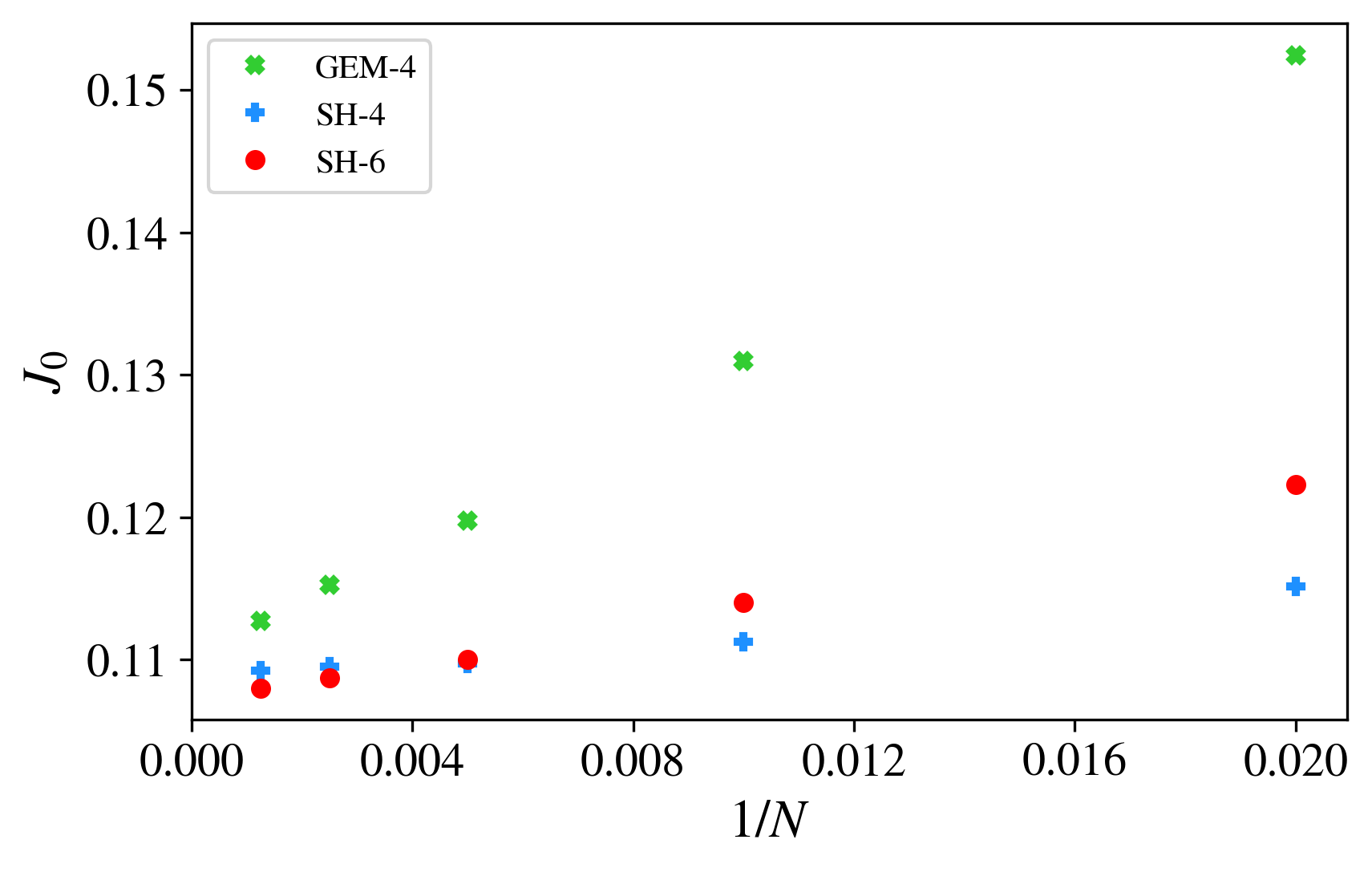}
  \caption{Results for $J_0$ as a function of the inverse number of soft particles, $1/N$.}
  \label{fig:scaling}
\end{figure}

As all the other observables involved in the mapping, $J_0$ naturally depends on the system size. 
A good final point consists in investigating the $J_0$ limit as the number of constituents of the 1D system increases. Simulations via SA of the soft system with growing number of particles $N$ yield the results reported in Fig.~\ref{fig:scaling}. The figure highlights how, for all the interaction potentials, $J_0$ as a function of $N$ tends to a finite asymptotic value typical of each model; this evidences that, for large $N$, the energy of the defect converges to a finite value (which corresponds to a classical metastable state) and $J_0$ becomes almost independent of the number of soft particles.

\section{Universality of pseudo--spin dynamics in 1D clustering}\label{sec:scaling}

In the previous sections we have established the emergence of effective Ising degrees of freedom via a nontrivial correspondence with a set of string variables in the soft particle systems under consideration. We observed that, for each model potential, a consistent $t$-dependent $J$ can be extracted for $t\lesssim 0.06$, which seems to converge to a $t\to 0$ value. We now show that the ratio $t/J$ is an approximate scaling variable for the pseudo--spin specific heat at temperatures $t\gtrsim 0.08 J$, provided a suitable $J$ is employed, which is not the low--temperature converged one. 

\begin{figure}[tbp]
  \includegraphics[width=\columnwidth]{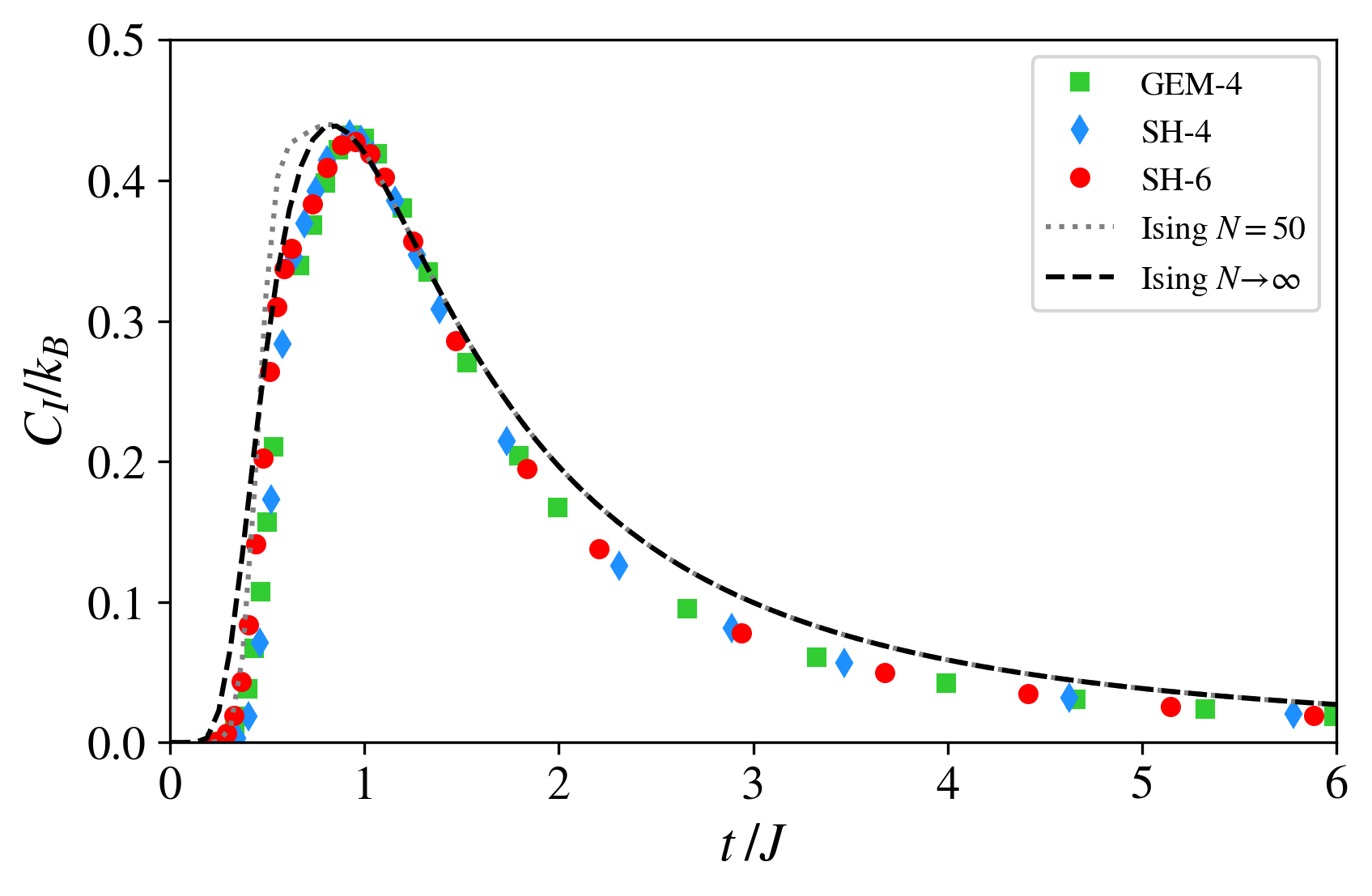}
  \caption{Specific heat in units of $k_B$ of the pseudo--spins for the considered model potentials, as a function of the scaling variable $t/J$, where $J$ is fitted from Fig.\ref{fig:ising_specific_heat} at the maxima of the specific heat. We also show the analytic results for the Ising model with $N_{\text{s}}=50$ spins in PBC and in the thermodynamic limit.}
  \label{fig:Cvising_scaled}
\end{figure}

In Fig.~\ref{fig:Cvising_scaled} we show the specific heat of the pseudo--spins in units of $k_B$, namely the data of Fig.~\ref{fig:ising_specific_heat} multiplied by $J^2$, as a function of $t/J$. The used values of $J$ are 0.15, 0.0865, and 0.136 for the GEM-4, SH--4 and SH--6 potentials, respectively, and correspond to the values fitted from the pseudo--spin specific heat at its maximum. We also plot the Ising specific heat in the thermodynamic limit and for $N_{\text{s}}=50$ in PBC (see Appendix). With the selected values of $J$, by construction the three sets of data cross the Ising curve at their peak: what is not obvious is that the Schottky-like anomaly appears at a common $t/J\simeq 1$, which is slightly above the value for the Ising model $t/J\simeq 0.83$, and therefore display the same peak magnitude $C_I\simeq 0.44k_B$, akin to the Ising model. The scaling is apparently valid for $t/J\gtrsim 0.08$. At higher temperatures, the results from the considered model potentials scale perfectly, independently of $J$, as can be appreciated in Fig.~\ref{fig:ising_specific_heat} where no multiplication by $J^2$ was performed. The reason is a high--temperature power-law decay of the specific heat with $1/t^2$. The coefficient in front of the $(J/t)^2$ behavior is however smaller than 1, which would hold for the Ising model. We speculate this scaling and its departure from the Ising model might be due to some coupling to the underlying phononic degrees of freedom. Conversely, discrepancies between the three sets of data can be seen at small temperature, which can be explained by both finite--size effects, which are known to be particularly large for the Ising model in PBC\cite{Lee_IsingPBC}, and by the fact that we used values of J fitted in a different temperature range.

\begin{figure}[tbp]
  \includegraphics[width=\columnwidth]{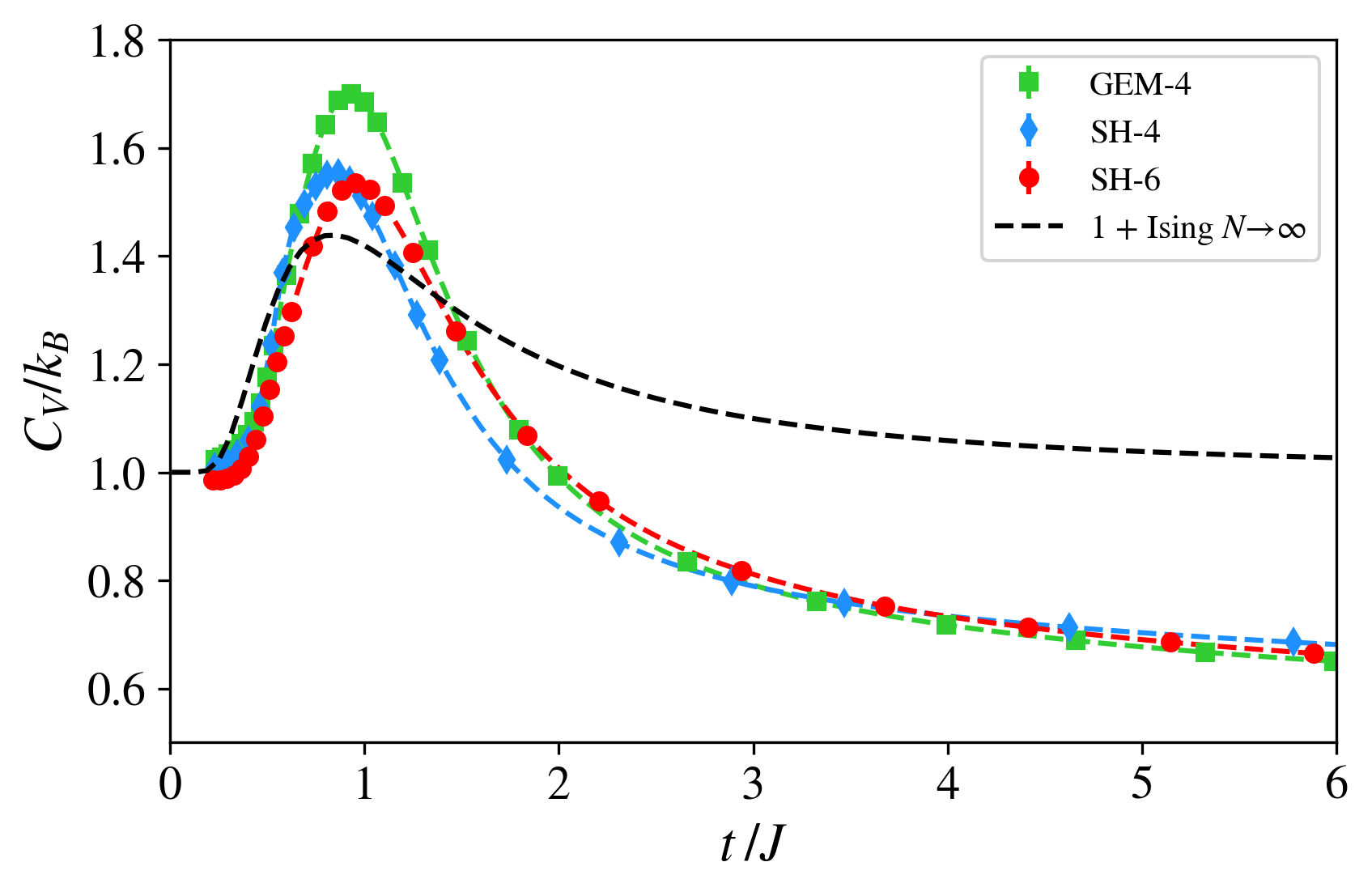}
  \caption{Specific heat in units of $k_B$ of the soft particles for the considered model potentials, as a function of the scaling variable $t/J$, where $J$ is fitted from Fig.\ref{fig:ising_specific_heat} at the maxima of the pseudo--spin specific heat. We also show the Ising model specific heat in the thermodynamic limit, shifted by 1.}
  \label{fig:Cvsoft_scaled}
\end{figure}

We also investigate whether there is a quantitative relation between the specific heat of the original soft systems, and the specific heat of the pseudo--spins. In fact, there is a striking similarity between Figs.~\ref{fig:cvsoft} and \ref{fig:ising_specific_heat}. In particular, the maxima of the specific heat appear at similar temperatures. In Fig.~\ref{fig:Cvsoft_scaled}, we thus plot again the data of Fig.~\ref{fig:cvsoft} as a function of the scaling variable $t/J$, with the same values of $J$ as reported in the previous paragraph. In the same figure, as a reference, we plot the Ising specific heat in the thermodynamic limit, $C_I/k_B=(J/t)^2/\cosh^2(J/t)$, shifted by 1, which, in the $t\to 0$ limit, accounts for the harmonic and kinetic contributions.

We observe that also for the clustering soft systems under consideration, the parameter $t/J\sim 1$, with the above specified values of $J$, marks the position of a peak in the specific heat. This is then a peculiar case of appearance of a Schottky anomaly in a classical non-magnetic system in the continuum. The magnitude of the anomaly varies and is in the range $0.5-0.7$ above 1. Interestingly, the high-temperature tails manifest a good scaling in $t/J$ for $t/J\gtrsim 3$. This scaling depends on the chosen $J$, since it is missing in Fig.~\ref{fig:cvsoft}. In fact, at variance with the pseudo--spin specific heats, the behavior in this temperature range is not consistent with a $t^{-2}$ power law.

\section{Conclusions}\label{sec:conclusions}

In this paper we have studied via Monte Carlo simulations the thermodynamic and structural properties
of 1D fluids of particles interacting by soft-core, repulsive pair potentials of the 
$Q^{\pm}$ class~\cite{ref_otto}, which allow spontaneous cluster formation. Three different functional
forms of the interaction were considered, and for each of them the number density was fixed at a value
commensurate with that of a dimer crystal. This investigation was prompted by a former study 
of a one-dimensional quantum boson fluid at zero temperature~\cite{ref_sei} where, on increasing
the strength of the inter-particle repulsion, a transition from a single-particle to a dimer Luttinger
liquid was observed. Remarkably, mapping this fluid into a magnetic system of pseudo--spins 
by the introduction of pseudo--spin string variables revealed that the spin-spin correlation function across the transition behaves like that of the quantum transverse Ising model. It then comes as a natural question, whether a similar situation may occur in the corresponding classical system. Importantly, in the latter case, the nature of the fluctuations is deeply different from the ones of the bosonic system; nonetheless, we have observed a rich phenomenology which is surprisingly similar to the one of the quantum fluid. 
In the classical case one expects that, on lowering the temperature, the fluid will eventually freeze into an ordered cluster crystal~\cite{likos:2007}. Unlike in the quantum fluid, the transition takes place for any value
of the repulsion strength, although in the one-dimensional system considered here, true long-range order
is still expected to develop only at vanishing temperature~\cite{prestipino:2015}. 

Similarly to the quantum case, we then introduced a mapping procedure onto string variables whereby the continuum, 
configurational degrees of freedom of the original system are replaced by Ising-like spins.
By construction, this mapping is such that a perfect or almost perfect dimer phase is turned 
into a ferromagnetic state, whereas a disordered configuration with no dimers corresponds to 
a paramagnetic state with random spin orientations. The remarkable point is that, in the low-temperature
limit, the pseudo--spins thus introduced, whose dynamics is dictated by that of the underlying fluid
of soft particles, do obey classical Ising statistics. Evidence of this was provided in several different
ways: 

i) The spin-spin correlation function is perfectly described by the corresponding expression
of the classical one-dimensional Ising model. As the temperature is lowered, the correlations decay more
and more slowly, pointing to the establishment of long-range order at zero temperature. Other thermodynamic quantities 
of the pseudo--spins, such as the internal energy, specific heat, and magnetic susceptibility, 
can also be described by their Ising expressions. In all cases, the spin-spin coupling constant $J$ 
obtained by fitting those expressions to the simulation results is found to depend on temperature. This is 
not surprising, in the light of the fact that the introduction of the string variables implies 
some kind of average over the configurational degrees of freedom (e.g., phonons), which may then result
in a state-dependent effective $J$. Nevertheless, at low temperature the values of $J$ obtained 
by this procedure become independent of the specific quantity under consideration, and approach 
a finite common limit as the temperature tends to zero. 

ii) At a given temperature, the probability of occurrence
of the first excited pseudo--spin state with respect to the ground state can be determined 
from the ratio of the soft-particle configurations which are mapped into either state. 
If the pseudo--spins obey Ising statistics, the coupling constant $J$ can be extracted by relating 
this ratio to that of the Boltzmann weights of the Ising Hamiltonian. At low temperature, 
the values of $J$ thus obtained fully agree with those determined from direct fit of the thermodynamic
properties according to the procedure of point~i). 

iii) The value of $J$ at zero temperature was estimated by comparing the energy of the ground state of the 
soft-particle system to that of its metastable state of lowest energy, determined via simulated annealing. 
The result is consistent with the extrapolation to zero temperature of the curves obtained 
by the procedures of points~i) and~ii) above. 

Since the mapping of the configurations of the original soft-particle fluid into an assembly 
of Ising pseudo--spins is not trivial and necessarily implies a significant loss of degrees of freedom,
there is not an obvious relationship between the thermodynamic observables of the fluid 
and their magnetic counterparts. Nevertheless, we think that the latter capture the role played
by the discrete degrees of freedom of the system. In particular,
the peak in the fluid specific heat as a function of temperature is mirrored by that in the specific heat
of the pseudo--spins (the Schottky anomaly), and its position turns out to be nearly independent 
of the potential considered, provided the temperature is rescaled by the effective magnetic coupling 
constant $J$ at the temperature of the peak. In the future, it would be interesting to study
the temperature dependence of $J$ beyond the phenomenological level considered here in order to clarify 
how this quantity is affected by the configurational degrees of freedom of the fluid. 

Another potentially interesting development would consist in extending the present analysis to clustering
involving more than two particles, or to systems in dimension larger than one. The latter development 
would allow to study the phase transition to cluster crystals at finite temperature, thereby making it 
more easily accessible to numerical simulation. While it is by no means obvious to us how the mapping 
considered here could be generalized to higher dimensions, since its present formulation clearly hinges 
on cluster formation in a 1D system, it is quite likely that the extension to larger clusters in 1D involves 
considering Potts models with higher spin.


\section*{Acknowldegements}
We acknowledge the CINECA awards IscraB\_MEMETICO (2018) and IscraC\_SOFT-ONE (2018) for the availability of high performance computing resources and support. D.P. acknowledges financial support by
Universit\`a degli Studi di Milano under Project
PSR2019\rule{0.15cm}{0.4pt}DIP\!\rule{0.15cm}{0.4pt}008-Linea~2.

F.M. and S.M. contributed equally to this work.


\appendix*
\section{Ising formulae for finite number of spins}

We report here the exact formulae for the Ising thermodynamical properties discussed in the main text, analytically calculated for a finite number of spins $N_{\text{s}}$ in PBC. In this Appendix we again normalize energy and temperature to the coupling parameter $U$.

Starting from the simple expression for the first--neighbor Ising energy $E_{I}=-J\,\sum_{i=1}^{N_{\text{s}}} \sigma_{i} \sigma_{i+1}$ (where the Bohr magneton $\mu_B$ is set equal to 1, and PBC are assumed), the partition function of the system can be easily derived, and the transfer matrix as well. Calling the eigenvalues of the transfer matrix $\lambda_+$ and $\lambda_-$, the free energy reads $F=-t \log(\lambda_+^{N_{\text{s}}}+\lambda_-^{N_{\text{s}}})$.
Setting $\zeta=J/t$, all the thermodynamic properties can then be derived, such as the internal energy per spin in units of $J$:
\begin{align}
\frac{\langle E_I \rangle}{J N_{\text{s}}} =& - \tanh \zeta \Biggl( \dfrac{1 + \tanh^{N_{\text{s}}-2} \zeta}{1+\tanh^{N_{\text{s}}} \zeta} \Biggr)\;,
\end{align}
the specific heat $C_I$ in units of $k_B J^2$:
\begin{align}
\frac{C_I}{k_B J^2} =&\frac{1}{t^2\cosh^2\zeta} \Biggl[1 + \frac{\tanh^{N_\text{s}} \zeta \csch^2 \zeta}{(1 + \tanh^{N_\text{s}}\zeta)^2} \times \Biggr.\\\nonumber
\Biggl.&\times\left( {N_\text{s}} - (1 + \tanh^{N_\text{s}} \zeta)\cosh(2 \zeta) \right)\Biggr]
\end{align}
the spin--spin correlation function:
\begin{equation}
g^\sigma_i=\frac{\sum_j\langle \sigma_j \sigma_{j+i} \rangle}{N_{\text{s}}}=\tanh ^{i}\zeta \Biggl(\frac{1+\tanh ^{N_{\text{s}}-2 i}\zeta}{1+\tanh ^{N_{\text{s}}}\zeta}\Biggr)\;,
\end{equation}
and finally the magnetic susceptibility:
\begin{equation}
\chi_I=\frac{e^{2\zeta}}{t} \Biggl(\frac{1-\tanh^{N_{\text{s}}} \zeta}{1+\tanh^{N_{\text{s}}} \zeta}\Biggr)\;.
\end{equation}

\end{document}